\documentclass[journal]{IEEEtran}
\usepackage{color}

\usepackage{lineno,hyperref,multirow,diagbox,multicol,balance,epstopdf,amsmath}
\usepackage{amsthm,amssymb}
\usepackage{mathrsfs}

\usepackage{tabularx}
\usepackage{multirow}
\usepackage{booktabs}
\newcommand{\topcaption}{%
\setlength{\abovecaptionskip}{0pt}%
\setlength{\belowcaptionskip}{10pt}%
\caption}   

\usepackage{cite}

\ifCLASSINFOpdf
   \usepackage[pdftex]{graphicx}
\else
\fi
 \usepackage[caption=false,font=footnotesize]{subfig}
\begin{document}
%
\title{MsEmoTTS: Multi-scale emotion transfer, prediction, and control for emotional speech synthesis}
%
%
%

\author{Yi~Lei, 
        Shan~Yang,
        Xinsheng~Wang,~\IEEEmembership{Student Member,~IEEE,}
        Lei~Xie,~\IEEEmembership{Senior Member,~IEEE }

\thanks{Corresponding author: Lei Xie}
\thanks{Yi Lei and Lei Xie are with the Audio, Speech and Langauge Processing Group, ASGO, School of Computer Science, Northwestern Polytechnical University, Xi’an 710072, China. Email: leiyi@npu-aslp.org, lxie@nwpu.edu.cn}
\thanks{Shan Yang is with Tencent AI Lab, Beijing 100086, China. Email: shaanyang@tencent.com}
\thanks{Xinsheng Wang is with the School of Software Engineering, Xi’an Jiaotong University, Xi’an 710049, China, and also with the School of Computer Science, Northwestern Polytechnical University, Xi’an 710072, China. Email: wangxinsheng@stu.xjtu.edu.cn}
}

%
%


\maketitle

\begin{abstract}
Expressive synthetic speech is essential for many human-computer interaction and audio broadcast scenarios, and thus synthesizing expressive speech has attracted much attention in recent years. Previous methods performed the expressive speech synthesis either with explicit labels or with a fixed-length style embedding extracted from reference audio, both of which can only learn an average style and thus ignores the multi-scale nature of speech prosody. In this paper, we propose MsEmoTTS, a multi-scale emotional speech synthesis framework, to model the emotion from different levels. Specifically, the proposed method is a typical attention-based sequence-to-sequence model and with proposed three modules, including global-level emotion presenting module (GM), utterance-level emotion presenting module (UM), and local-level emotion presenting module (LM), to model the global emotion category, utterance-level emotion variation, and syllable-level emotion strength, respectively. In addition to modeling the emotion from different levels, the proposed method also allows us to synthesize emotional speech in different ways, i.e., transferring the emotion from reference audio, predicting the emotion from input text, and controlling the emotion strength manually. Extensive experiments conducted on a Chinese emotional speech corpus demonstrate that the proposed method outperforms the compared reference audio-based and text-based emotional speech synthesis methods on the emotion transfer speech synthesis and text-based emotion prediction speech synthesis respectively. Besides, the experiments also show that the proposed method can control the emotion expressions flexibly. Detailed analysis shows the effectiveness of each module and the good design of the proposed method.
\end{abstract}


%
\IEEEpeerreviewmaketitle

\section{Introduction}
%
%
%
%
\IEEEPARstart{T}{ext} to speech (TTS) aims at generating human-like speech from input texts. The traditional TTS systems are mainly based on the statistical parametric methods~\cite{black2007statistical, ze2013statistical, ling2013modeling, qian2014training, watts2016hmms}. In recent years, the success of the neural sequence-to-sequence (seq2seq) paradigm brings significant improvement to the TTS performance~\cite{wang2017tacotron, shen2018natural, li2019neural, sotelo2017char2wav, ren2019fastspeech, yu2020durian, yang2020localness, chen2020adaspeech}, making it possible to synthesize natural speech, and thus the speech interaction has been popular in many virtual assistants, like Siri, Alexa, Google Assistant, etc. However, synthetic speech that only sounds natural is insufficient to meet the requirement of immersive human-computer interaction and audio broadcast applications, in which producing expressive speech with emotions is essential. In this paper, we will focus on the emotional speech synthesis.

A straightforward way to perform the expressive speech synthesis is to train a TTS model with a manually labeled database according to the style categories, e.g., emotion categories, in which the style label works as the conditional information to guide the TTS model to produce speech with the same style category~\cite{lee2017emotional, lorenzo2018investigating}. However, these methods can only learn an average representation for each style, making it impossible to produce diverse speech styles according to different textual contents from the same style category. This is obviously different from our human beings' ability that we can convey the same speech style with various subtle variants. To address this issue, style transfer TTS has been a popular strategy in recent years~\cite{skerry2018towards, wang2018style, hu2020unsupervised, kenter2019chive, hsu2018hierarchical, li2021controllable}. In the style transfer TTS, reference audio with the expected speech style is used to provide the reference speech style, and the aim is to synthesize speech with the same style as that of reference audio. Obtaining the style-discriminative embedding from reference audio is crucial for the emotion transfer TTS task. In~\cite{skerry2018towards}, a reference encoder was proposed to extract a fixed-length embedding to model the prosody of the reference signal. To learn the acoustic expressions, GST model~\cite{wang2018style} is proposed with the soft interpretable ``Global Style Tokens'', which shows good performance on learning style representations. 

While the reference audio-based expressive TTS method can synthesize more diverse speech than those label-based methods, the style embedding extracted from reference audio still can only present a global style. The learned style embedding from the reference signal can not convey speech style from multiple levels within an utterance.

In fact, the style expressions of human speech are multi-scale in nature~\cite{selkirk1986derived, liberman1977stress, beckman1986intonational}, which varies from coarse to fine granularity, rather than mono-scale. From the perspective of coarse granularity, we can summarize the overall style or emotion category of an entire utterance, for instance, we can say that the emotion conveyed by an utterance is ``happy". From the view of fine granularity, each pronunciation unit in a sentence has its own characteristics such as intensity, speed, pitch, and energy~\cite{li2021towards}. In addition to the above granularities, some prosodic expressions like the fluctuations of intonation are also reflected in between- and cross-pronunciation unit relationships~\cite{tseng2005fluent}. This multi-scale property in prosody results in plenty of styles with rich and subtle emotion changes in human speech. Therefore, it is insufficient to model speech style from a single aspect. To sufficiently model the speech emotion in different levels, in this paper, we propose MsEmoTTS, a \textit{multi-scale} model that models emotions from different levels for emotional speech synthesis.

Our proposed multi-scale emotional speech synthesis model is based on the typical attention-based sequence-to-sequence architecture and with three proposed modules, i.e., global-level emotion presenting module (GM), utterance-level emotion presenting module (UM), and local-level emotion presenting module (LM). 
To be specific, 1) GM is concerned with what the emotion category is conveyed by the entire utterance, such as happy or sad; 2) UM focuses on the prosody pattern (e.g., intonation) within an utterance, which varies within the utterance; 3) LM aims at providing emotion strengths for speech pronunciation units, e.g., syllables and phonemes, which can control the intensity of localized emotional expressions. In all of these modules, the emotion information can either be extracted from reference audio or predicted from the input text, which makes that the proposed model can realize the emotional speech synthesis by either transferring the emotion from reference audio or predicting from the input text. Moreover, we can also manually provide the global emotion category in GM, and define the local emotion strengths in LM, which allows us to control the synthetic emotional speech as expected. Extensive experiments have shown that the proposed MsEmoTTS achieves good performance on synthesizing emotional speech in different ways.

Our preliminary work has been presented in~\cite{lei2021fine}, in which only global-level and local-level emotion information are considered. In the current work, the preliminary model is extended with UM to model the utterance-level variance, which shows the effect on improving emotional speech synthesizing. Moreover, the modules for modeling the global emotion category and local emotion strengths are substantially improved in the current work, and this improvement has been proved in the component analysis. Overall, the main contributions of this paper are summarized as follows:
\begin{itemize}
  \item A multi-scale emotional speech synthesis model is proposed, which consists of three sub-modules, i.e., GM, UM, and LM, to provide the global-level emotion category, utterance-level emotion variation, and local-level emotion strengths, respectively.
  \item The proposed model is a unified and flexible model that allows us to synthesize emotional speech in different ways, including transferring the emotion from reference audio, predicting the emotion from the input text, and obtaining the emotion from the manual definition. All of these are very practical.
  \item Extensive experiments show the good performance of the proposed method on synthesizing emotional speech in different ways. The detailed component analysis demonstrates the effectiveness of emotion information delivery from each level and indicates the good design of the proposed method.
\end{itemize}

The rest of the paper is organized as follows. Section~\ref{Related works} introduces the related work. Section~\ref{Model architecture} describes the proposed method. Section~\ref{Experiments and Discussions} introduces the details of experiments, and the corresponding results are presented in Section~\ref{sc:exp_results}. The component analysis is shown in Section~\ref{exp_hierarchical}. Section~\ref{sc:discussion} discusses the performance and limitation of the current work, and also the possible future research. Finally, the paper is concluded in Section~\ref{Conclusion}.

\section{Related Work}
\label{Related works}
This section reviews related work on expressive speech synthesis based on explicit representation and reference audio. We also review related studies on predicting speaking styles from input text for synthesizing expressive speech.

\subsection{Explicit label based expressive speech synthesis}
\label{Explicit labels-based expressive speech synthesis}
With an expressive speech corpus that is labeled with various styles, it is intuitively to train an expressive TTS model with the style (e.g. emotion and speaker) category as the conditioning information ~\cite{lorenzo2018investigating, luong2017adapting, liu2021controllable}. Lee et al.~\cite{lee2017emotional} proposed an emotional end-to-end neural speech synthesizer, injecting the emotion label vector to the attention-based decoder. Luong et al.~\cite{luong2017adapting} took labels of multiple aspects, i.e. gender, age, and speaker, into consideration as conditioning information. With these explicit labels, it can perform adaptation to novel speakers and control the synthetic speech output in the form of labels input.

In addition to the explicit label of emotion category, there are also some recent approaches aiming at generating expressive speech with controllable emotion rendering~\cite{rabiee2019adjusting, zhu2019controlling}. With continuous variables, the emotion strength of synthetic speech can be controlled flexibly to generate expressions of different emotion intensities.

Similar to those methods that take the style category label to synthesize speech, we also use a category embedding from a lookup table as the global emotion information to our TTS system. In addition to the given emotion category, the proposed method can also predict the emotion from input text and model emotional speech in different granularities.

\subsection{Reference audio based expressive speech synthesis}
\label{Reference audio-based expressive speech synthesis}
The reference audio based TTS is to synthesize speech with the style that transferred from reference audio. A straightforward way is to obtain the style representation of reference audio, which then can work as the conditional information to guide the speech synthesizing~\cite{watts2015sentence, skerry2018towards,wang2018style, zhang2019learning}. In~\cite{skerry2018towards}, a reference encoder trained with an unsupervised method is proposed to obtain the prosody embedding (representation) from reference audio. Then, the prosody embedding is fed into the autoregressive decoder of Tacotron architecture, together with speaker information and transcript encoder output. The Global style token (GST) model~\cite{wang2018style} is an updated method to learn the style representation by encoding various speaking styles from reference audio into a fixed number of tokens, which are then combined to obtain the final style representation. To improve the interpretability of the learned style tokens, in~\cite{wu2019end}, the number of style tokens is particularly set to the number of emotion categories, and a classification loss in terms of emotion category is further introduced to bring an explicit relation between style tokens and emotion categories. Cai et al.~\cite{cai2021emotion} enables a GST-based model to generate expected emotions by jointly training an emotion predictor, with the help of a pre-trained cross-domain speech emotion recognition model. Um et al.~\cite{um2020emotional} proposed an inter-to-intra emotional distance ratio algorithm that specifically minimizes the intra-cluster embedding vectors and maximizes the inter-cluster ones, and introduces an effective interpolation that can conduct emotion intensity control for the synthetic speech.

Instead of learning the style embedding for the whole utterance, some recent research focused on fine-grained speech expressions~\cite{tan2020fine, lee2019robust, klimkov2019fine, zhang2020learning}. Tan et al.~\cite{tan2020fine} proposed a fine-grained model to extract style embeddings from phoneme-level speech segments using collaborative learning and adversarial learning strategies. In~\cite{klimkov2019fine}, phoneme-level aggregated prosodic features are also adopted in a seq2seq-based TTS system to improve the robustness of prosody transfer. 

Recently, some efforts have been conducted to model the style in multiple scales or hierarchy~\cite{an2019learning,li2021towards, chien2021hierarchical}. In~\cite{an2019learning}, the authors extended the GST to a hierarchical GST architecture, in which several GST layers are used with a residual connection, to learn hierarchical embedding information implicitly. In their model, the first GST layer performs well at the speaker discrimination, while the representations from deeper GST layers tend to contain finer speaking styles or emotion variations. Different from this implicit way, in~\cite{li2021towards}, a reference encoder-based model is trained explicitly to extract phoneme-level and global-level style features from mel-spectrograms. In this paper, the proposed method also learns information from different scales explicitly. In addition to the local level and global level, the utterance level for extracting utterance intonation is also considered in our model. Besides, the proposed model can not only transfer the emotion from reference audio but also predict the emotion from the input text. Furthermore, clearer interpretable control is also supported by the current work.

\subsection{Predict speaking style from input text}
\label{Prediction speaking style from input text}
Compared to label-based and reference audio-based methods, directly predicting the speaking style from input text is more practical and flexible, as it allows the TTS system to bypass the dependency on the manually selected labels or reference audio during the inference stage~\cite{stanton2018predicting, zhang2021extracting}. In the Text-Predicted Global Style Token (TP-GST) proposed in ~\cite{stanton2018predicting}, the authors extended the GST by predicting the style rendering from text only, making it possible to get the stylistic speech automatically rather than using explicit labels or reference audio. Instead of predicting global styles, some recent work tried to predict the styles from a fine-grained level, such as phoneme level ~\cite{tan2020fine} or word level \cite{zhang2021extracting}. 

Different from those methods that either only predict the global style information or only focus on the local style information, MsEmoTTS proposed in this paper takes different levels into consideration and integrates emotion transfer, prediction, and control in a unified model.

\section{Multi-scale emotional expression modeling}
\label{Model architecture}
\begin{figure*}[ht]
        \centering
        \includegraphics[width=1.0\linewidth]{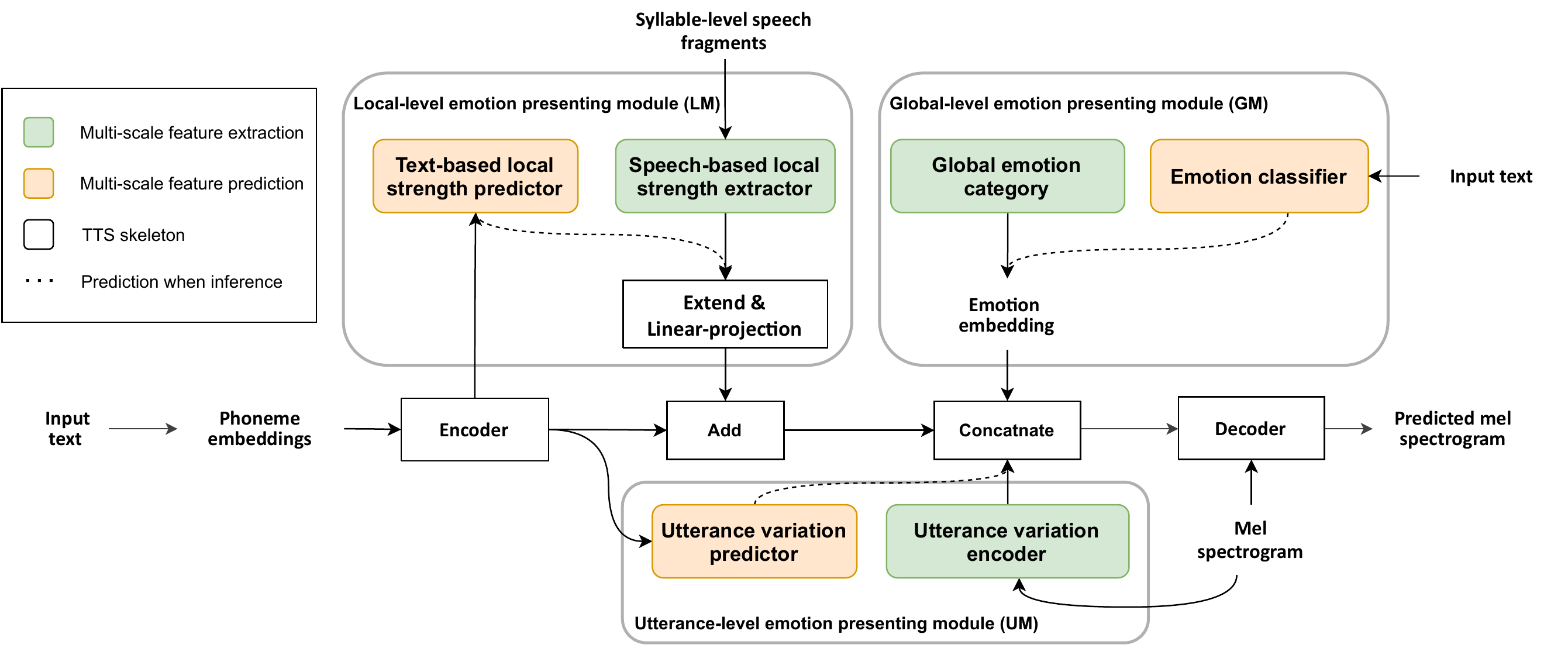}
        \caption{The architecture of MsEmoTTS. The green boxes are the procedure of multi-scale emotion feature extraction, orange boxes are the procedure of multi-scale feature prediction, and white boxes are the basic acoustic model's skeleton. The emotion classifier in GM and speech-based local strength extractor in LM are pre-trained, and their parameters will not be updated during the training of the TTS system. }
         \vspace{-10pt}
        \label{fig:system}
\end{figure*}

The framework of MsEmoTTS is shown in Fig~\ref{fig:system}. As shown in this figure, the backbone of the proposed method follows the attention-based auto-regressive acoustic model in~\cite{shen2018natural}. To model the emotion from different levels, three modules, i.e., global-level emotion presenting module (GM), utterance-level emotion presenting module (UM), and local-level emotion presenting module (LM), are proposed to model the emotion from the global level, utterance level, and local level, respectively.

In this section, we will first give an overview of MsEmoTTS, following which the details of each module will be introduced. Furthermore, the way to realize the emotion transfer, prediction, and manual control for emotional TTS with the proposed method will also be introduced in this section.

\subsection{Overview}
\label{sc:overview}

The basic skeleton of the acoustic model is based on the Tacotron~\cite{wang2017tacotron} and Tacotron2~\cite{shen2018natural}. Specifically, the encoder part is the same as that of Tacotron~\cite{wang2017tacotron}, and the decoder part is the same as that of Tacotron2~\cite{shen2018natural}. As for the attention mechanism, instead of the LSA attention mechanism adopted in \cite{wang2017tacotron,shen2018natural}, the GMM-based attention mechanism ~\cite{battenberg2020location}, which is more robust for long sequences, is adopted in our work. In addition to this skeleton, the proposed three modules, i.e., GM, UM, and LM, allow the model to obtain 1) the overall emotion category; 2) the subtle variation in the intonation of the emotion within an utterance, and 3) the emotion strength of each speech unit, e.g., syllable in this work. During the training stage, the training data is organized as \textit{text-audio-emotion} triads. We denote a triad of training data as $\{X,Y,e\}$, where $X$ is input textual sentence presented as a sequence of phonemes, i.e., $X=\{x_1,x_2,..,x_T\}$, with sequence length of $T$. $Y$ is the speech representation, i.e., mel-spectrograms, and $e$ is the emotion category. Assuming the local emotion strength of each speech unit is known. 
With the extension of syllable-level strengths, the local emotion strength sequence can be presented with the same length as the phoneme sequence, i.e., $S=\{s_1,s_2,...,s_T \}$. The hidden emotion representations of global level, utterance level, and local level can be obtained by the following functions respectively:
\begin{equation}
\label{eq:overall}
\begin{array}{l}
\begin{aligned}
  & {h_{global}} = {f_1}\left( e \right)  \\ 
  & {h_{utt}} = {f_2}\left( Y \right)  \\
  & {h_{local}} = {f_3}\left( S \right), 
  \end{aligned}
\end{array}
\end{equation}
where $f_1$ is an emotion category embedding function, $f_2$ is the utterance variation encoder, and $f_3$ is a linear projection function. The goal of training process is to reconstruct $Y$ with the input textual sequence $X$ and conditional multi-scale emotion representations, including ${h_{global}}$, ${h_{utt}}$, and ${h_{local}}$.

Details of each module to obtain emotion representations from a specific level will be introduced in the following subsections. The final objective function to train this model will also be introduced in Section~\ref{sc:loss_fuction}. 

\subsection{Global-level emotion presenting module (GM)}
\label{sc:method_GM}
The goal of GM contains two aspects. One is to provide the conditional emotion embedding during the training process, in which the emotion category is provided by the training set. At this training stage, the global emotion embedding is obtained with a trainable look-up table with embeddings for different emotion categories, which means that all utterances from the same emotion category share the same emotion embedding. 

The other goal of GM is to predict the emotion category from input text during the inference stage. To this end, a pre-trained text-based emotion classifier is adopted. 
The emotion classifier is based on a pre-trained BERT-Base~\cite{kenton2019bert} Chinese model released by Google~\footnote{The pre-trained BERT-Base model can be found at \url{https://github.com/google-research/bert}}. To perform the emotion classification task, a linear layer and a softmax function are added following the backbone of the BERT. Then the pre-trained BERT is fine-tuned along with the added linear layer in the textual emotion classification task.

The output of the softmax layer is the possibility that the input text belongs to each emotion category. Normally, we can obtain the specific emotion category based on the maximum probability of the softmax output. Then, an emotion embedding can be chosen based on this predicted emotion category~\cite{lei2021fine, lee2017emotional}, and the chosen emotion embedding is called \textit{hard} global emotion embedding. However, compared with the speech emotion classification, the emotion classification with the text as input is more challenging. The main reason is that the correspondences between the text and emotion are not uniquely determined, e.g., the same text could be read with different emotions for different intentions, which introduces some ineluctable misclassifications by the text-based emotion classifier. These misclassifications could lead to disharmonious emotion expressions for the speech synthesizing. To address this problem, instead of using the \textit{hard} embedding, here, a \textit{soft} emotion embedding is proposed to represent the global emotion. 

Different from the fixed and finite \textit{hard} global emotion embeddings, the proposed \textit{soft} global embedding is a weighted emotional embedding based on the possibility of each emotion category calculated by the text-based emotion classifier (the output of the softmax layer). Specifically, assuming there are $M$ different emotion categories $E=\{e_1,e_2,...e_M\}$, the \textit{soft} global emotion embedding is obtained by:

\begin{equation}
\label{eq-global}
  h_{global} = \sum_{i}^M (p_{i} \cdot {f_1}(e_{i})),
\end{equation}
where $p_i$ is the possibility out from the softmax layer of the emotion classifier for predicting the input text to emotion $e_i$. Note that the $h_{global}$ calculated in Eq.~\ref{eq:overall} is for the training process, in which $e$ is the ground-truth emotion category, while here the $h_{global}$ is for the inference stage. In addition to this automatically learned \textit{soft} global emotion embedding, we can also manually specify the emotion category to realize the control of the global emotion.

\subsection{Utterance-level emotion presenting module (UM)}
\label{sc:UM}
The utterance-level emotion presenting module (UM) is proposed to obtain the utterance-level emotion variation, defined as the emotional fluctuation of speech intonation over time in this paper. Similar to GM, the goal of UM also has two aspects: 1) provide the utterance emotion variations with speech as input during the training stage, and 2) predict the utterance emotion variations with the text as input during the inference stage. Specifically, during the training stage, as explained in Eq.~\ref{eq:overall}, the representation of the utterance-level emotion variation is obtained by an utterance variation encoder with ground-truth mel-spectrograms as input. 

With the aim to achieve the ability on predicting the utterance-level emotion variations from text, a variation predictor with the text encoder outputs as input is optimized during the training stage, which is supervised by $h_{utt}$ output from the utterance variation encoder. We denote the output of this text-based variation predictor as $\hat{h}_{utt}$. The optimization of this variation predictor is achieved by minimizing the loss function:
\begin{equation}
   {\cal L}_{utt} = \left\| {{h_{{\text{utt}}}} - {{\hat h}_{{\text{utt}}}}} \right\|_2^2.
\end{equation}

Both the utterance variation encoder and the text-based variation predictor are 1d convolution-based models. Specifically, the utterance variation encoder consists of two 1d convolution layers followed by layer normalization and dropout. Mean pooling at the time axis is applied on the output to obtain a vector representation for the utterance emotion variation. The text-based variation predictor consists of a 1d convolution layer with the ReLU activation function and layer normalization, and two fully connected layers. The input for the predictor is the linguistic feature learned from the text encoder, and the output is the predicted utterance variation.

\subsection{Local-level emotion presenting module (LM)}
Similar to GM and UM, LM is to: 1) provide the speech unit-level, e.g., syllable-level, emotion strength with speech as input during the training stage, and 2) predict the speech unit-level emotion strength with the text as input during the inference stage. We utilize a pre-trained speech-based local strength extractor and a text-based local strength predictor to realize the two goals. In addition to predicting the local emotion strength from the text, during the inference stage, we can also directly define the emotion strength manually to realize the strength control. Details of the \textit{speech-based local strength extractor}, \textit{text-based local strength predictor}, and \textit{local emotion strength control} will be introduced in the following.

\textbf{Speech-based local strength extractor}.
Intuitively, a straightforward way to get a speech-based local strength extractor is to train a strength prediction model with a speech corpus labeled with local strengths. Unfortunately, such a corpus is usually unavailable since it is difficult to manually label the fine-grained speech strength due to the complex emotion expression. Instead, a ranking-based method named relative attributes~\cite{parikh2011relative} is adopted in this work to learn the strengths in an unsupervised manner, where the emotion strength is regarded as an attribute of speech.

Specifically, assuming there are two speech sets $N$ and $E$, in which the training speech $I = \{i\}$ is represented in $\mathbb{R}^n$ by emotion-related acoustic features $\{\boldsymbol{f_{i}}\}$. In the training set $N$, the acoustic features are extracted from neutral speech, while features in $E$ are from a kind of emotional speech such as ``happy''. In light of the consideration that an emotional speech utterance should always have higher emotion strength than that of neutral speech, our goal is to learn a ranking function
\begin{equation}
\label{eq-local}
r(\boldsymbol{f_{i}}) = \boldsymbol{w}\boldsymbol{f_{i}}
\end{equation}
for calculating the emotion strength of speech $i$, in which $\boldsymbol{w}$ should satisfy:

\begin{equation}
\label{eq-local}
\begin{array}{l}
\forall(\boldsymbol{f_{i}} \in E~and~\boldsymbol{f_{j}} \in N): \boldsymbol{w} \boldsymbol{f_{i}}>\boldsymbol{w} \boldsymbol{f_{j}} \\
 \forall((\boldsymbol{f_{i}},\boldsymbol{f_{j}}) \in E~or~(\boldsymbol{f_{i}},\boldsymbol{f_{j}}) \in N): \boldsymbol{w} \boldsymbol{f_{i}}=\boldsymbol{w} \boldsymbol{f_{j}}.
\end{array}
\end{equation}

Following Parikh et al.~\cite{parikh2011relative}, a Newton's method~\cite{chapelle2007training} is adopted to optimize the weighting vector $\boldsymbol{w}$:
\begin{equation}
\begin{aligned}
\operatorname{minimize} &\left(\frac{1}{2}\left\|\boldsymbol{w}^{T}\right\|_{2}^{2}+C\left(\sum \xi_{i j}^{2}+\sum \gamma_{i j}^{2}\right)\right) \\
 \text { s.t. } & \boldsymbol{w}^{T}\left(\boldsymbol{f_{i}}-\boldsymbol{f_{j}}\right) \geq 1-\xi_{i j} ; \forall(\boldsymbol{f_{i}} \in E~and~\boldsymbol{f_{j}} \in N) \\
 &\left| \boldsymbol{w}^{T}\left(\boldsymbol{f_{i}}-\boldsymbol{f_{j}}\right)\right| \leq \gamma_{i j} ; \forall((\boldsymbol{f_{i}},\boldsymbol{f_{j}}) \in E~or~(\boldsymbol{f_{i}},\boldsymbol{f_{j}}) \in N) \\
 & \xi_{i j} \geq 0 ; \gamma_{i j} \geq 0,
\end{aligned}
\end{equation}
where $C$ is utilized to control the trade-off between the margin and the size of the slack variables $\xi_{i j}$ and $\gamma_{i j}$.

\begin{figure}[ht]
        \centering
        \includegraphics[width=1.0\linewidth]{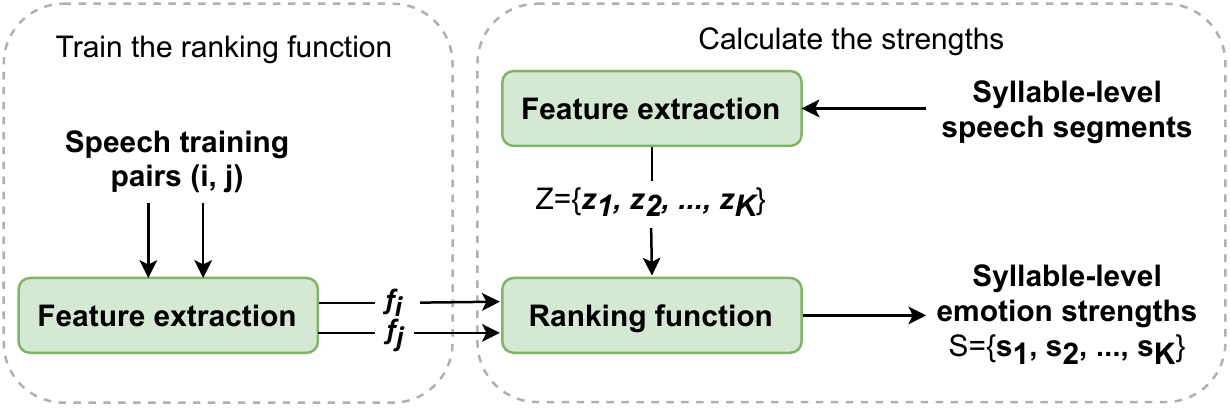}
        \caption{The procedure of calculating syllable-level emotion strengths.}
         \vspace{-2pt}
        \label{fig:rankingfunction}
\end{figure}

In practice, we use utterance-level speech to train the above ranking function and take the syllable-level speech unit to obtain the local emotion strengths as shown in Fig~\ref{fig:rankingfunction}. This is because Chinese is a character-based language and each character is naturally pronounced as a syllable. The speech unit boundaries are obtained by an HMM-based alignment model. Considering that emotional strength has a strong relationship with the pitch and energy, the acoustic features for calculating strength is extracted by the openSMILE~\cite{eyben2010opensmile}, resulting in a 384-dimensional emotion-related feature. Given a sequence of speech fragment features $Z=\{\boldsymbol{z_1},\boldsymbol{z_2},...,\boldsymbol{z_K}\}$, the corresponding emotion strength sequence  $S=\{s_1,s_2,...,s_K\}$ could be obtained through
 \begin{equation}
    s_i = \boldsymbol{w}\boldsymbol{z_i}.
 \end{equation}

Then, representations of the local emotion strengths, referred to as $h_{local}$, can be achieved by a linear transformation layer. For the convenience of manual control during the inference stage, the local emotion strengths are normalized into $[0,1]$.

\textbf{Text-based local strength predictor}. 
To predict the local emotion strength from the input text during the inference stage, a text-based strength predictor is jointly trained along with the acoustic model. Similar to the training of text-based variation predictor in Section~\ref{sc:UM}, this text-based local strength predictor is also trained under the supervision provided by the pre-trained speech-based local strength extractor. The output of the text-based local strength predictor is denoted as ${\hat S}$, and the training of the text-based local strength predictor is to minimize the loss function:
\begin{equation}
   {\cal L}_{local} = \left\| {S - {\hat S}} \right\|_2^2.
\end{equation}
The network of the local strength predictor is the same as the utterance variation encoder.

\textbf{Local emotion strength control}. As described above, the local emotion strengths are calculated by the speech-based local strength extractor. For each emotion category, we normalize all strengths into rational numbers in the range of $[0, 1]$. As the assumption in the local extractor of Eq.~\ref{eq-local}, higher local strength means stronger emotion intensity expression. During the training stage, all the local strengths are utilized as conditioning of the acoustic model, and the text-based local strength predictor is jointly trained with it. During the inference stage of emotion strength control, the strength can be manually assigned as any rational value in the range of $[0, 1]$ as needed. It also can be appointed as the strengths from the reference signal. In this way, fine-grained local strength control can be conducted by the manual instructions or reference speech.

\subsection{Objective function}
\label{sc:loss_fuction}
The loss function of the acoustic model backbone, referred to as ${\cal L}_{acoustic}$, follows that in Tactoron2~\cite{shen2018natural}, which is defined as the mean squared error (MSE) between the synthesized mel-spectrograms and the ground-truth mel-spectrograms. The total loss function is defined as
\begin{equation}
   {\cal L} =  {\cal L}_{acoustic} + \lambda_1 {\cal L}_{local} + \lambda_2 {\cal L}_{utt},
\end{equation}
where $\lambda_1$ and $\lambda_2$ are hyper-parameters to balance the weights of different losses. Note that ${\cal L}_{local}$ and ${\cal L}_{utt}$ are only used to optimize the text-based local strength predictor and text-based variation predictor respectively. 

\subsection{Inference}
During the inference stage, it allows us to synthesize emotional speech in different ways in the unified model, i.e., 1) transfer the emotion from reference audio, 2) manually control the emotion, and 3) predict the emotion from input text.

\textbf{Emotional speech synthesis by transferring the emotion from reference audio}. With the speech-based local strength extractor and speech-based utterance variation encoder, it is easy to perform the emotion transfer TTS. In this task, the global emotion category, local emotion strength, and utterance variation are provided by reference audio, and the goal is to make the synthesized speech imitate the emotion expressions of the reference audio. Note that in non-parallel emotion transfer, the lengths of reference and input text are different, so we conduct linear interpolation on the reference strength sequence to obtain the target local strengths.

\textbf{Emotional speech synthesis by manual control}. In the proposed method, we can manually define the global emotion category in GM and provide the local strengths in LM as expected, making it possible to customize arbitrary emotion expressions. Due to that the latent utterance-level variation is hard to control, only the global emotion category and local strengths are manually controlled, and the utterance variation can be provided by the prior mean of the training data. For the local strength control, all speech units can be assigned to any strength scale in $[0, 1]$, which allows to enhance or weaken some units in specific positions, or gradually increase or decrease the emotion expressions if needed.

\textbf{Emotional speech synthesis by predicting the emotion from input text}. With the text-based emotion classifier, text-based utterance variation predictor, and text-based local strength predictor, the proposed model can automatically learn the multi-scale emotion representations from the input text and synthesize emotional speech that is consistent with the content, in which way the emotional speech synthesis can bypass the dependency on reference audio and manual labels.

\section{Experiments}
\label{Experiments and Discussions}
To evaluate the performance of MsEmoTTS on the emotional speech synthesis task with various conditions, i.e., based on reference audio, manual control, and based on input text only, extensive experiments are performed on a Chinese emotional speech corpus. In this section, the databases for the emotional speech synthesis and also for the training of sub-modules will be introduced. Besides, implementation details, compared methods, and the evaluation method will also be introduced. 

\subsection{Database}
\textbf{Database for emotional speech synthesis}. An emotional speech corpus is adopted for training our TTS system. This corpus was recorded by a professional female voice actor, which contains neutral speech and six kinds of emotional speech, including happiness, anger, sadness, surprise, fear, and disgust. There are 10000 utterances, about 10 hours audio, in the neutral data, and 2000 utterances, about 2 hours audio, in each emotion category. This emotional speech corpus is suitable for our emotional TTS task because of the strong emotion expressions and distinct styles of different emotions. In addition to the training stage, experiments on emotion transfer and control are also carried on this corpus. To this end, 100 utterances of each emotion category are randomly selected for the test.

\textbf{Database for the emotion classifier}. In GM, the emotion classifier, which is based on the pre-trained BERT, is fine-tuned for the emotion classification task. In order to improve the generalization of the classification and avoid overfitting, we adopt extra emotional text data from Chinese Weibo texts in NLPCC2013 and NLPCC2014\footnote{Data can be found at \url{http://tcci.ccf.org.cn/conference/2013/pages/page04_tdata.html} and \url{http://tcci.ccf.org.cn/conference/2014/pages/page04_sam.html}} to train this emotion classifier. In NLPCC datasets, there are about 3500 happy, 2600 angry, 3300 sad, 1100 surprise, 400 fear, 4100 disgust and 38000 neutral text sentences respectively. During the inference stage, we use the text data from another internal emotional speech corpus to verify the rationality and ability of generalization in emotion prediction from input text only. There are 200 sentences for each emotion category in our test data. All our experiments on emotion prediction utilize this test data.

\subsection{Implementation details }
\label{model-details}

Speech is represented by 80-band mel-scale spectrograms with the frame shift of 20ms. To reconstruct waveforms from the predicted mel-spectrograms, a multi-band WaveRNN~\cite{kalchbrenner2018efficient}, which is a 4-band WaveRNN with 512 RNN units, is adopted to perform this reconstruction. For the fair comparison, the WaveRNN is only trained with the real mel-spectrograms and speech pairs, and no fine-tuning process with the synthesized mel-spectrograms exists for any compared method.

\subsection{Compared methods}
As this work, to our knowledge, is for the first time that has the ability to synthesize emotional speech based on different emotion transfer, prediction, and control in a unified model, there is no direct work that can be compared directly and fairly. Therefore, different methods are compared for different tasks. Specifically, for the reference audio-based emotional speech synthesis, a state-of-the-art reference audio embedding method, named GST~\cite{wang2018style}, is adopted to compare with our method on the emotional speech synthesis by transferring the emotion from reference audio. For the text-based emotional speech synthesis, a variant of GST, named TPSE-GST~\cite{stanton2018predicting}, which is a GST-based model with the ability to predict style embeddings from text, is compared with our method on emotional speech synthesis by predicting the emotion from the input text. For the fair comparison, the same vocoder and Tacotron-based basic skeleton of acoustic model are utilized with all systems in experiments.

\subsection{Evaluation methods}

\textbf{Evaluation metric for objective evaluation}. In the reference audio-based emotional speech synthesis, the reference audio is practical with the different transcription from the input text, which is called non-parallel transfer. However, when the input text for TTS is the same as the transcription of the reference audio, which is called parallel transfer, we can easily evaluate the synthesized results objectively by comparing them with the reference audio that works as the ground truth.

In this work, to perform the objective comparison between our method and the compared method, a parallel transfer task is conducted in the reference-audio based emotional speech synthesis (see Section \ref{exp_transfer}). Here, mel-cepstral distortion (MCD) is utilized to calculate the difference between the synthesized results and ground-truth speech. The MCD score quantifies the distortion between MFCCs extracted from two audios, and the smaller value in our experiments means that the synthesized result is more similar to the ground truth, indicating better performance. In practice, dynamic time warping (DTW)~\cite{sakoe1978dynamic} is adopted to align the predicted and target spectrogram.

\textbf{Human perceptual rating experiment}. Except for the parallel transfer TTS, there is no ground-truth speech for other cases, such as non-parallel transfer TTS and predicting emotion from the text, which makes the evaluation of synthesized speech be a perceptual task. Here, mean opinion score (MOS), comparative mean opinion score (CMOS), and A/B preference test are conducted to evaluate the performance of emotion expressions. 

The MOS test is to ask participants to rate a given audio based on a specific rule. For instance, in Section~\ref{exp_prediction}, participants are asked to rate the audio based on how well this audio's emotion is consistent with the sentiment of the input text. In this paper, the score of MOS test ranges from $1$ to $5$ with the interval of $0.5$, in which $1$ means very bad and $5$ means excellent.

Both CMOS and A/B tests are preference experiments given two compared models. In the A/B test, participants are asked to choose the better one based on a specific rule. Besides, the \textit{Neutral} is also a possible option, which can be chosen when the participant thinks both demos are similar. Compared with the A/B test, the CMOS test~\cite{loizou2011speech} asks participants to rate for the comparative degree. The rating of CMOS is a rational number which means the quality of the second model compared to the first one. The rating score ranges from $-3$ to $3$, in which $-3$ means the first model is much better than the second one, while $3$ means the second one is much better than the first one, and $0$ means the compared two models are similar. A smaller absolute CMOS value means that the gap between the two models is smaller.

All the subjective rating experiments are held in a quiet studio room, and 30 Chinese native speakers using headphones participated. We particularly asked the listeners to ignore the audio quality and focus on the emotional expressions. In non-parallel emotion transfer, we select 5 text transcripts from each emotion category. For each text transcript, we randomly select 5 reference signals per category from the other 5 emotion categories, resulting in 125 generated non-parallel speech for each emotion category and 750 synthesized audio in total. For the listening tests, we randomly choose 20 generated utterances per emotion category for a total of 120 generated utterances for each listener. In other experiments, 20 utterances are randomly selected for each emotion category.

\section{Experimental results}
\label{sc:exp_results}
Experiments on emotional speech synthesis tasks in different ways, including transferring the emotion from reference audio, predicting the emotion from the input text, and obtaining the emotion from the manual definition, are presented in this section.

\subsection{Emotional speech synthesis by transferring the emotion from reference audio}
\label{exp_transfer}

We first compare our MsEmoTTS with the GST model on the parallel emotion transfer task, which allows us to compare these two methods objectively. The results are presented in Table~\ref{tab:mcd-pt}. As can be seen, in this objective evaluation, the proposed MsEmoTTS model significantly outperforms the GST model. Specifically, the MCD of MsEmoTTS is 11.5\% relatively lower than that achieved by the GST model, which means modeling the real emotional speech more closely, demonstrating the superiority of the MsEmoTTS method on the parallel emotion transfer speech synthesis task. 

\begin{table}[htb]
\centering
\topcaption{Objective comparison with the GST model on the parallel emotion transfer speech synthesis task. The performance is evaluated with MCD. Lower MCD means better performance.}
\label{tab:mcd-pt}
\setlength{\tabcolsep}{12mm}
\begin{tabular}{cc}
\toprule
Methods & MCD (dB) \\ \midrule
GST    & 4.10 \\ 
MsEmoTTS (Proposed)    & \textbf{3.63} \\ \bottomrule
\end{tabular}

\end{table}

\begin{table*}[tp]
\centering
\topcaption{Subjective comparison with the GST model on both parallel and non-parallel emotion transfer speech synthesis tasks. The MsEmoTTS is the proposed method, and $p$ denotes the $p$-value of a $t$-test between two models. CMOS and A/B preference tests are performed according to the emotion similarity. A positive CMOS value means that the MsEmoTTS method is better than the GST model, while a negative CMOS value means the GST model is better.}
\label{tab:cmos-transfer}
\setlength{\tabcolsep}{3mm}
\begin{tabular}{c|c|cccc|c|c|cccc}
\toprule
    \multicolumn{1}{c|}{\multirow{2}{*}{Parallel}}&
    \multicolumn{1}{c|}{\multirow{2}{*}{CMOS}} &
    \multicolumn{4}{c|}{Preference(\%)} &
    \multicolumn{1}{c|}{\multirow{2}{*}{Non-parallel}} &
    \multicolumn{1}{c|}{\multirow{2}{*}{CMOS}} &
    \multicolumn{4}{c}{Preference(\%)} \\
    \cline{3-6} \cline{9-12}
    & & GST & Neutral & MsEmoTTS  & $p$ & & & GST & Neutral & MsEmoTTS & $p$  \\ \midrule
 Happiness  & 0.329 & 17.9 & 35.7 & \textbf{46.4} & 0.023 & Happiness  & 0.156 & 26.2 & 34.5 & \textbf{39.3} & 0.043 \\
    Anger  & 0.093 & 21.4 & \textbf{39.8} & 38.8 &  0.157 & Anger  & 0.287 & 22.8 & 34.7 & \textbf{42.5} & 0.037 \\
    Sadness  & 0.757 & 14.3 & 21.4 & \textbf{64.3} &  0.014 & Sadness  & 0.367 & 20.7 & 31.5 & \textbf{47.8} &0.026 \\
    Surprise  & 0.843 & 10.6 & 13.7 & \textbf{75.7} &  $<$ 0.001 & Surprise  & 0.517 & 20.5 & 15.8 & \textbf{63.7} & $<$ 0.001 \\
    Fear  & 0.153 & 10.7 & 43.4 & \textbf{45.9} & 0.038 & Fear  & 0.083 & 23.7 & \textbf{39.4} & 36.9 & 0.245 \\
    Disgust  & 0.924 & 10.3 & 14.0 & \textbf{75.7} &  $<$ 0.001 & Disgust  & 0.641 & 9.8 & 13.0 & \textbf{77.2} & $<$ 0.001 \\ \midrule
    Average   &  0.520&  14.2&  27.7& \textbf{58.1} &  0.018 & Average   &  0.342 & 20.6  & 28.2  & \textbf{51.2} & 0.021  \\ \bottomrule
\end{tabular}
\end{table*}

The human perceptual evaluation results for the methods both on the parallel emotion transfer task and non-parallel emotion transfer task are shown in Table~\ref{tab:cmos-transfer}, in which CMOS and A/B preference test results are reported. In the CMOS test, participants are asked to rate a pair of audios synthesized by the GST model and our MsEmoTTS in terms of the emotion similarity with the reference audio. 
As results from this table, in terms of all emotion categories, all CMOS scores for both parallel emotion transfer task and non-parallel emotion transfer task are greater than 0, indicating the better performance of MsEmoTTS on synthesizing emotional speech by transferring emotions from the reference audio. It is worth noting that almost all CMOS scores, except for that in terms of \textit{Anger} in the parallel emotion transfer task and \textit{Fear} in the non-parallel emotion transfer task, are larger than 0.1, which demonstrates the significant superiority of MsEmoTTS on the emotion transfer task compared with the GST model.

The A/B test gives further insight into the comparison of the baseline GST model and the proposed MsEmoTTS. In this test, the option \textit{Neutral} could be chosen if participants cannot choose a superior result from the compared samples. As shown in this Table, the average preference scores of MsEmoTTS outperform the GST model with a big margin in both parallel and non-parallel emotion transfer speech synthesis tasks. Except for the parallel transfer task with the emotion of ''anger" and the non-parallel transfer task with the emotion of ``fear", the proposed method significantly outperforms the GST model on all other cases ($p$-value smaller than 0.05). All those results demonstrate that the proposed method can convey emotion information from the reference audio effectively, and achieves overall superiority on the reference audio based emotional speech synthesis compared with the GST model.

\subsection{Emotional speech synthesis by predicting the emotion from input text}
\label{exp_prediction}

\begin{table}[t]
\vspace{10pt}
\centering
\topcaption{Subjective comparison with TPSE-GST model on the text-based emotional speech synthesis task for emotion naturalness. MOS results are reported with 95\% confidence interval. Higher MOS means better performance. ``Recordings" means the ground-truth samples that provide an upper boundary.}
\label{tab:mos}
\scalebox{1.0}{
\setlength{\tabcolsep}{5.5mm}
\begin{tabular}{c|c|c}
\toprule
MOS  & TPSE-GST & MsEmoTTS (Proposed) \\ \midrule
Happiness  & 3.68 $\pm$ 0.153 & \textbf{4.19 $\pm$ 0.084} \\
Anger      & 3.15 $\pm$ 0.195 & \textbf{4.04 $\pm$ 0.128} \\
Sadness    & 3.60 $\pm$ 0.162 & \textbf{3.95 $\pm$ 0.156} \\
Surprise  & 3.14 $\pm$ 0.158 & \textbf{4.15 $\pm$ 0.117} \\
Fear       & 3.41 $\pm$ 0.169 & \textbf{4.05 $\pm$ 0.075}  \\
Disgust    & 3.45 $\pm$ 0.135 & \textbf{3.75 $\pm$ 0.156} \\ \midrule
Average    & 3.405 $\pm$ 0.162 & \textbf{4.02 $\pm$ 0.119} \\ \bottomrule
\end{tabular}
}
\end{table}

With the trained text-based emotion classifier in GM, text-based variation predictor in UM, and text-based local strength predictor in LM, the proposed method can be easily used to synthesize emotional speech based on input text only, bypassing the dependency on the reference audio. The comparison results between MsEmoTTS and TPSE-GST on the text-based emotional speech synthesis are shown in Table~\ref{tab:mos}.

The MOS in this table is obtained in a human perceptual rating test. In this test, participants are asked to rate a given sample based on how well the emotion of this sample (audio) is consistent with the corresponding text. The higher MOS means the emotion of speech has better consistency with the sentiment of the transcription. The bold in this table means the better performance between TPSE-GST and MsEmoTTS. As shown in this table, MsEmoTTS outperforms the TPSE-GST in terms of all emotion categories. Specifically, the average MOS achieved by our method is 18.1\% relatively higher than that of TPSE-GST, indicating the better performance of MsEmoTTS on the text-based emotional speech synthesis compared with TPSE-GST, which demonstrate that MsEmoTTS is suitable for emotional speech synthesis by predicting the emotion from the input text. 

\subsection{Emotional speech synthesis by manual control}
\label{exp_control}

\begin{figure} 
    \centering
	  \subfloat[happiness]{
       \includegraphics[width=0.45\linewidth]{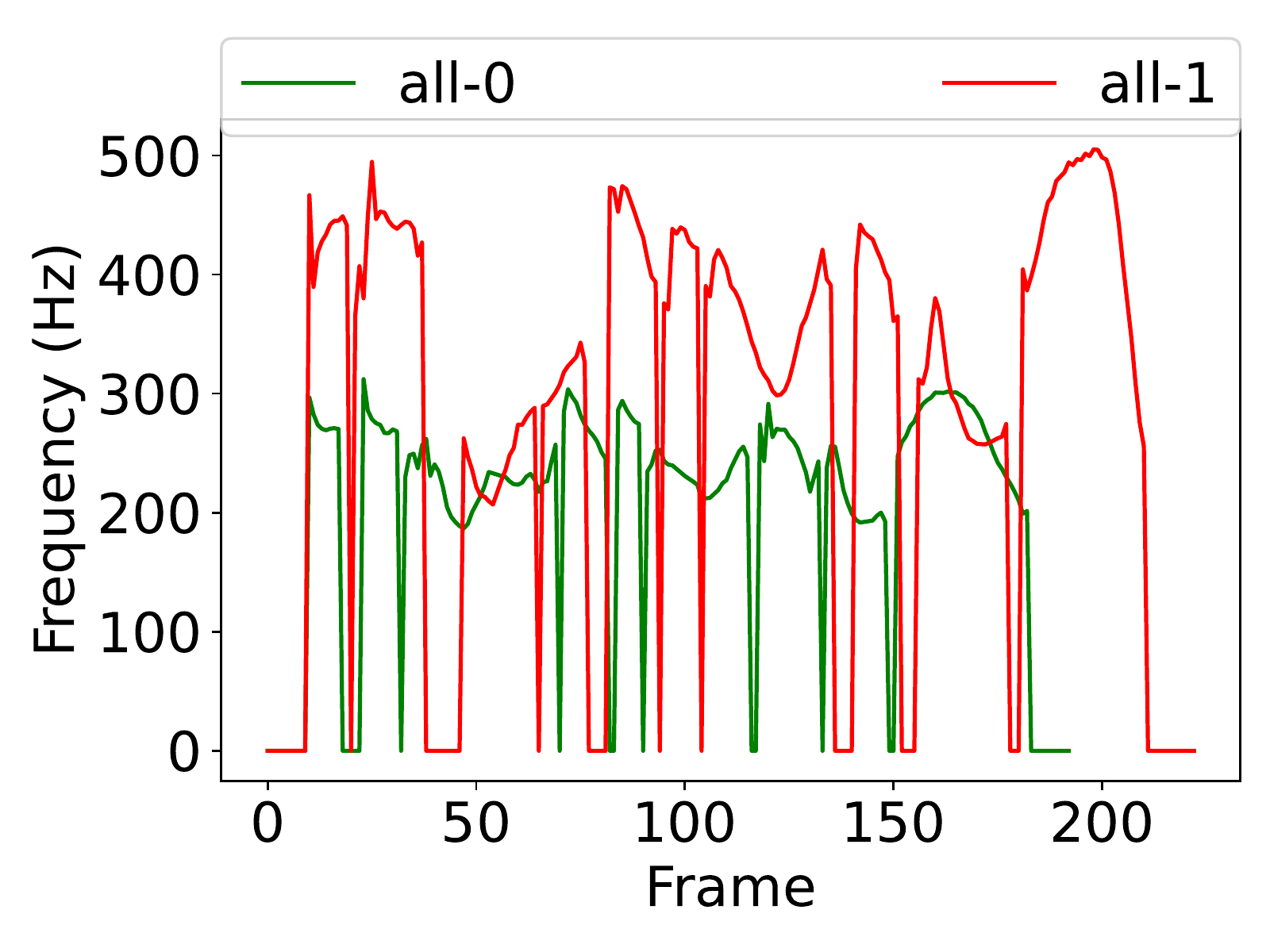}}
    \label{1a}\hfill
	  \subfloat[anger]{
        \includegraphics[width=0.45\linewidth]{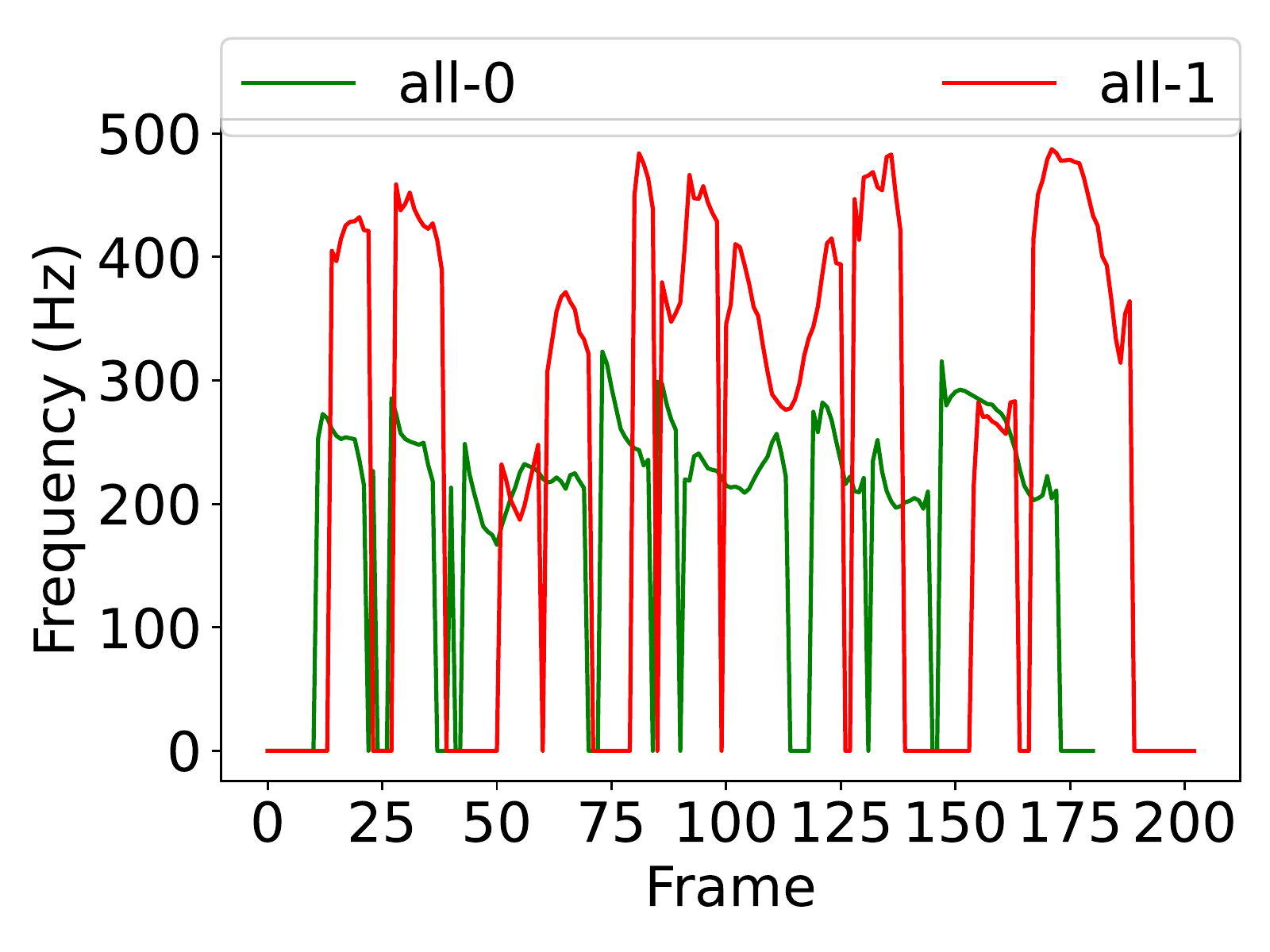}}
    \label{1b}\\
	  \subfloat[sadness]{
        \includegraphics[width=0.45\linewidth]{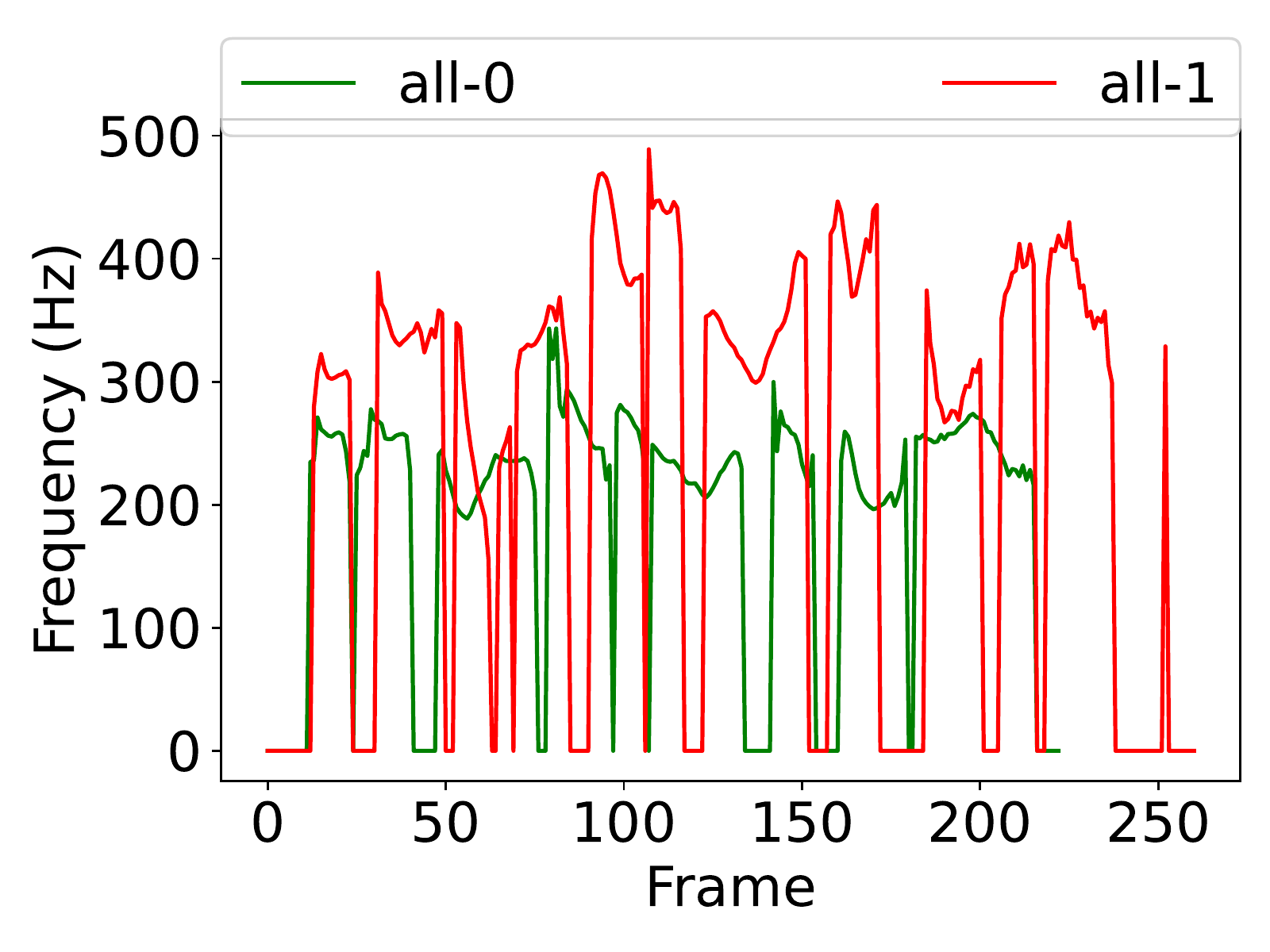}}
    \label{1c}\hfill
	  \subfloat[surprise]{
        \includegraphics[width=0.45\linewidth]{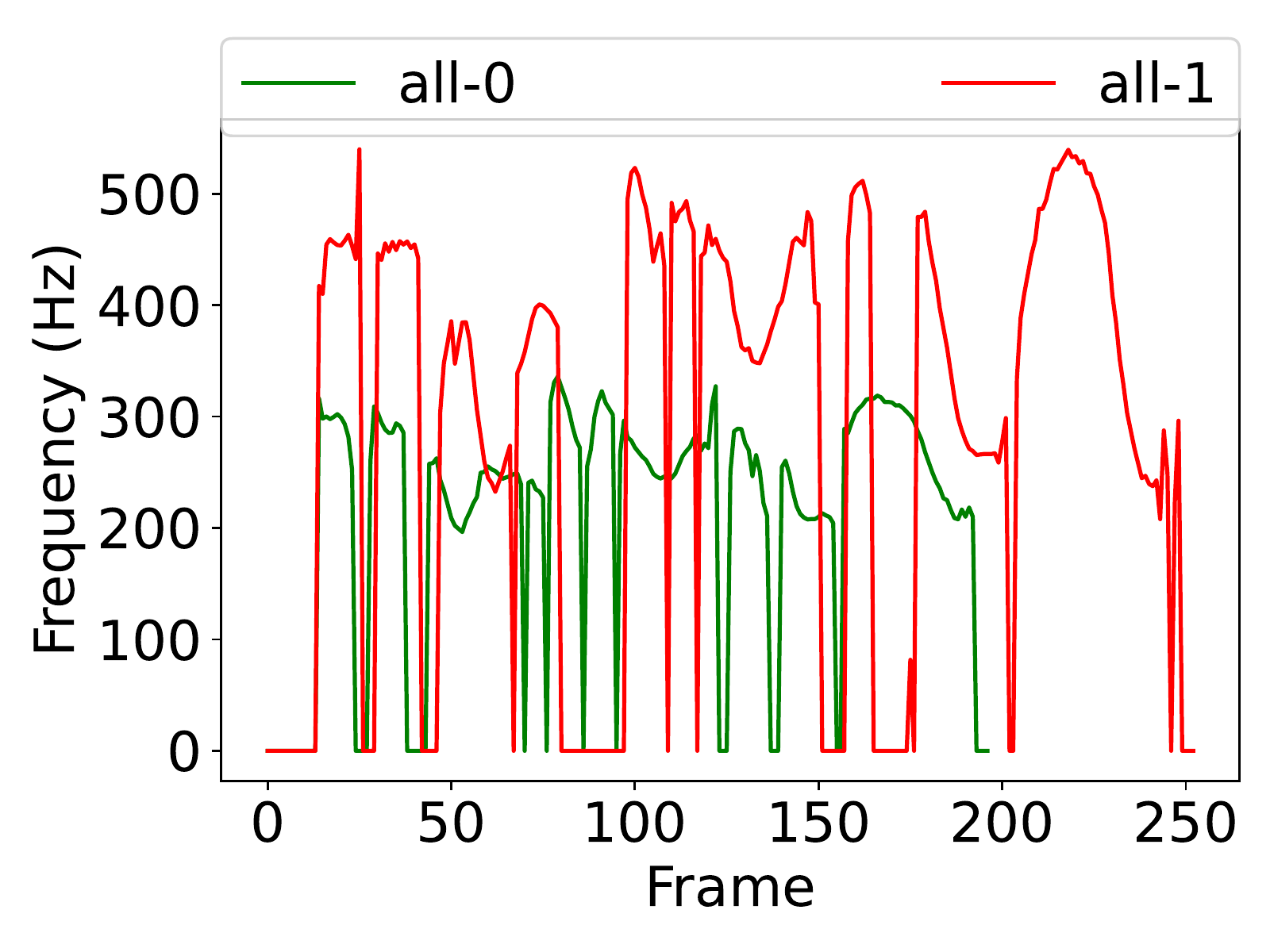}}
     \label{1d} 
     \subfloat[fear]{
        \includegraphics[width=0.45\linewidth]{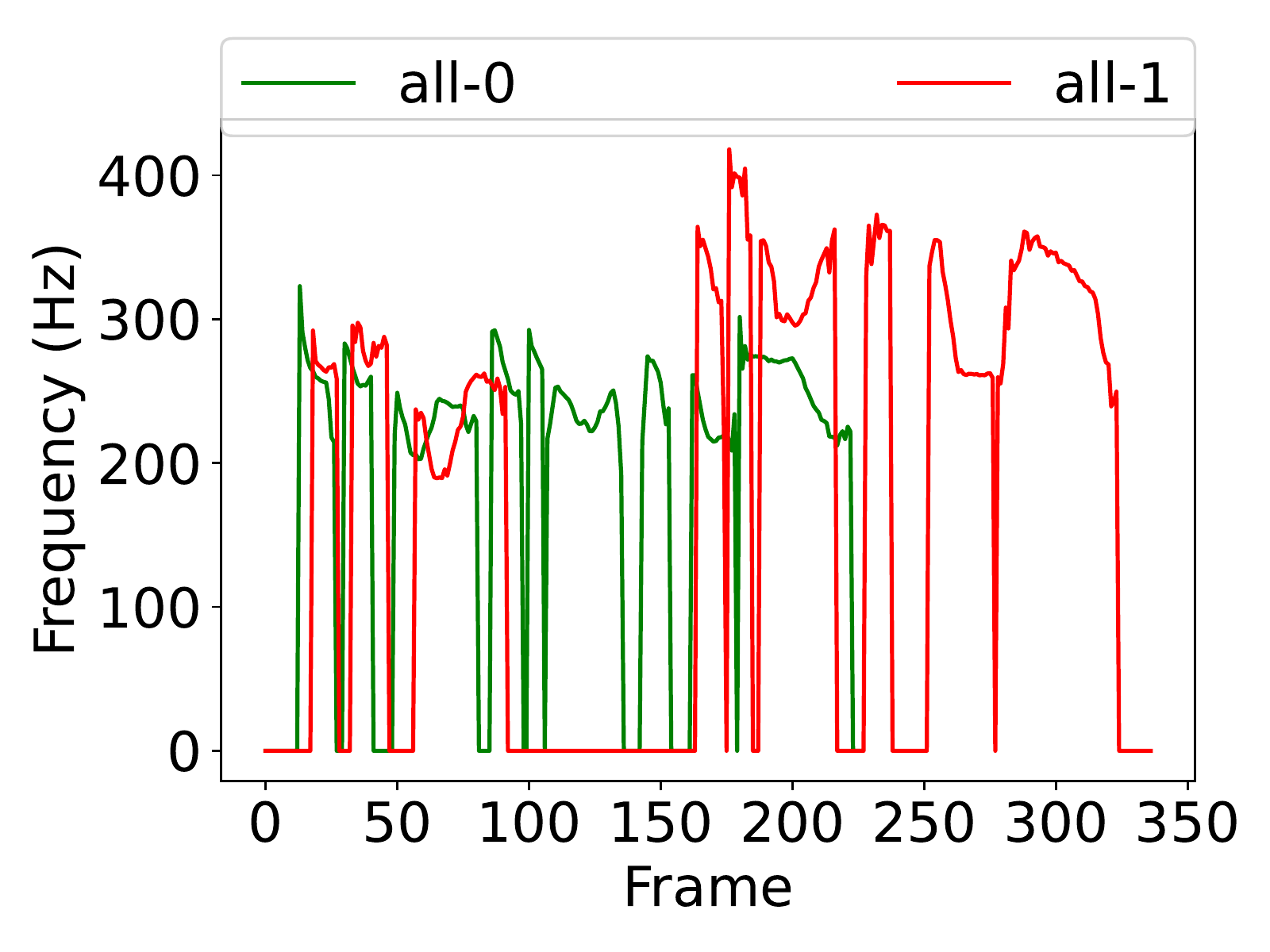}}
    \label{1e}\hfill
	  \subfloat[disgust]{
        \includegraphics[width=0.45\linewidth]{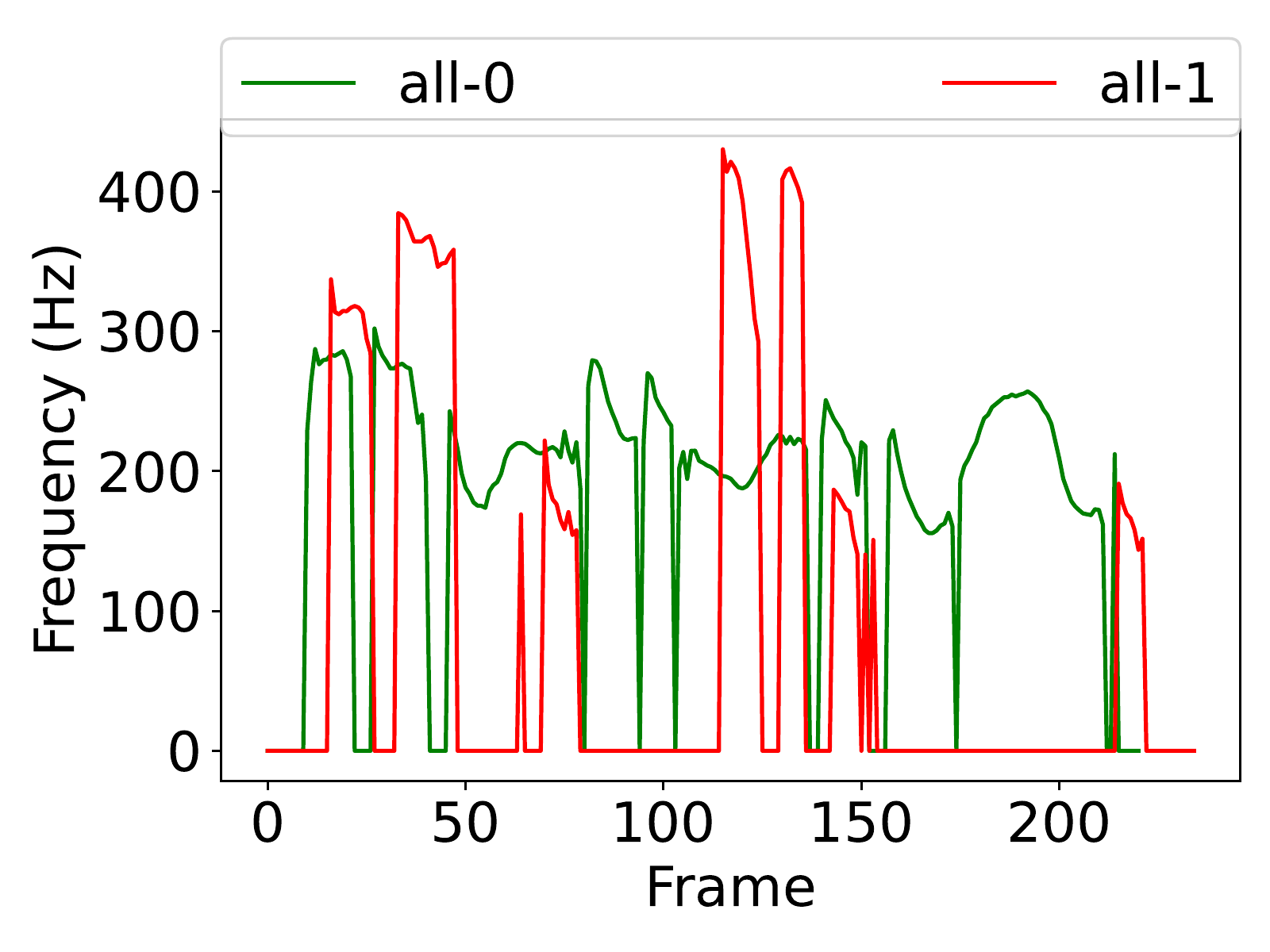}}
     \label{1f} 
	  \caption{F0 curves of synthesized samples with fixed strengths of different emotion strengths. ``All-0'' means that all syllable strengths are set as 0 for synthetic speech, while ``all-1'' means that all syllable strengths are set as 1.}
	  \label{fig:ctl_01} 
\end{figure}

With the local strengths in LM and emotion category in GM, MsEmoTTS can be manually controlled to synthesize emotional speech. In this section, we present the ability of the proposed method to synthesize emotional speech with manual control. To this end, synthesized speech with the same input text but different local emotion strengths and global emotion categories is produced by our method. For the intuitive observation of pitch variation, F0 curves of the synthesized speech with different strengths are drawn in the same figure as shown in Fig.~\ref{fig:ctl_01}. For each emotion category, we assign the strengths of all syllables to be 0 or 1. It can be seen that for emotion categories of ``happiness'', ``anger'', ``sadness'', and ``surprise'', higher local strengths always lead to higher F0 and longer duration, which express stronger emotional intensity. For the ``fear'' emotion, the duration of synthesized speech with strength 1 is much longer than that with strength 0. This result is in line with our intuitive understanding that we express more fear speech at a lower speaking speed and longer pauses. For the ``disgust'' emotion, the synthesized speech with strength 1 has abrupt changes in F0, which means the higher strength may lead to obvious emotional changes represented as longer pauses and shorter voiced speech. The results for all emotions indicate that the emotion strength of synthetic speech can be manually controlled successfully.

\begin{figure} 
    \centering
	  \subfloat[happiness]{
       \includegraphics[width=0.45\linewidth]{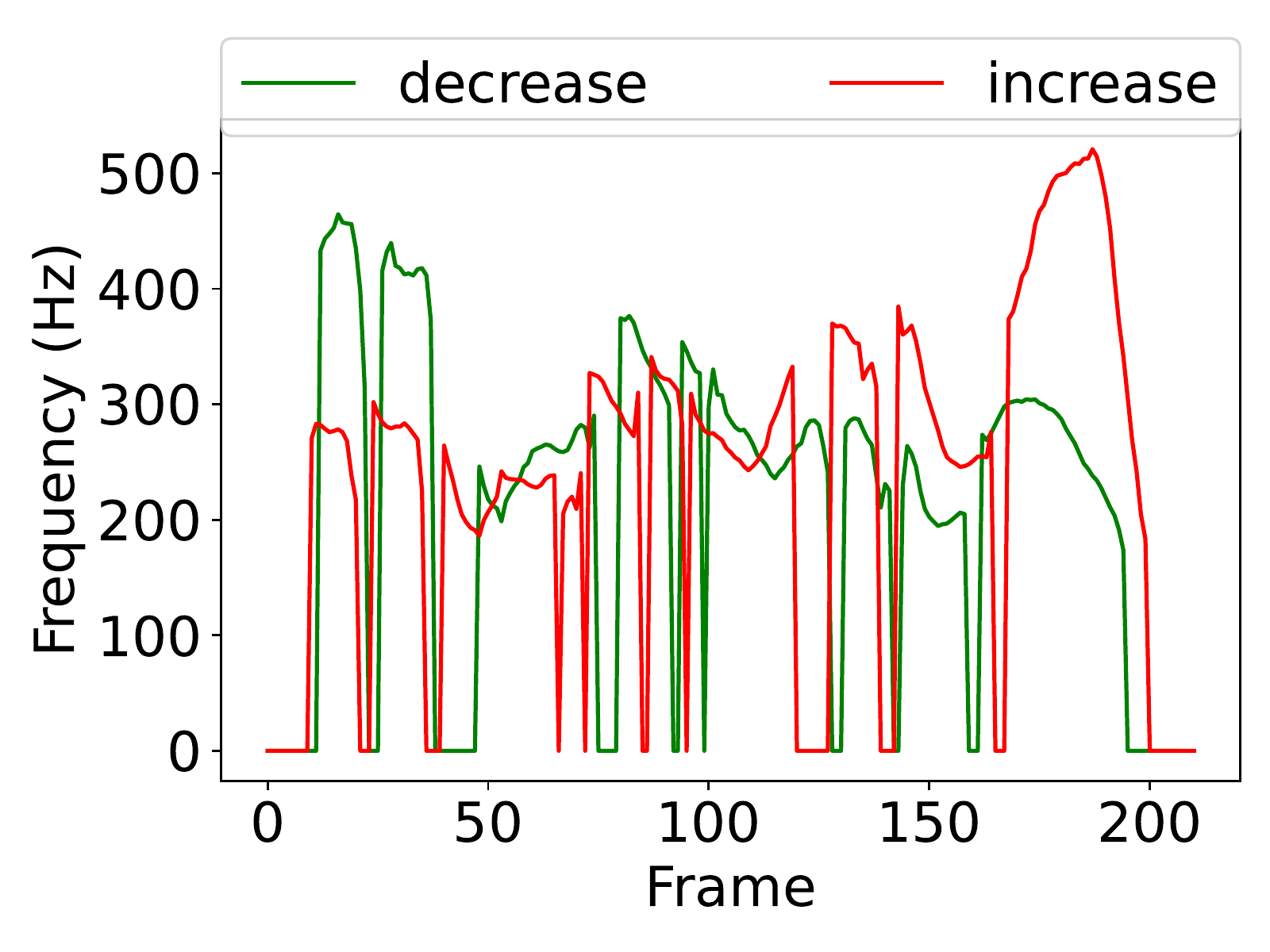}}
    \label{1a}\hfill
	  \subfloat[anger]{
        \includegraphics[width=0.45\linewidth]{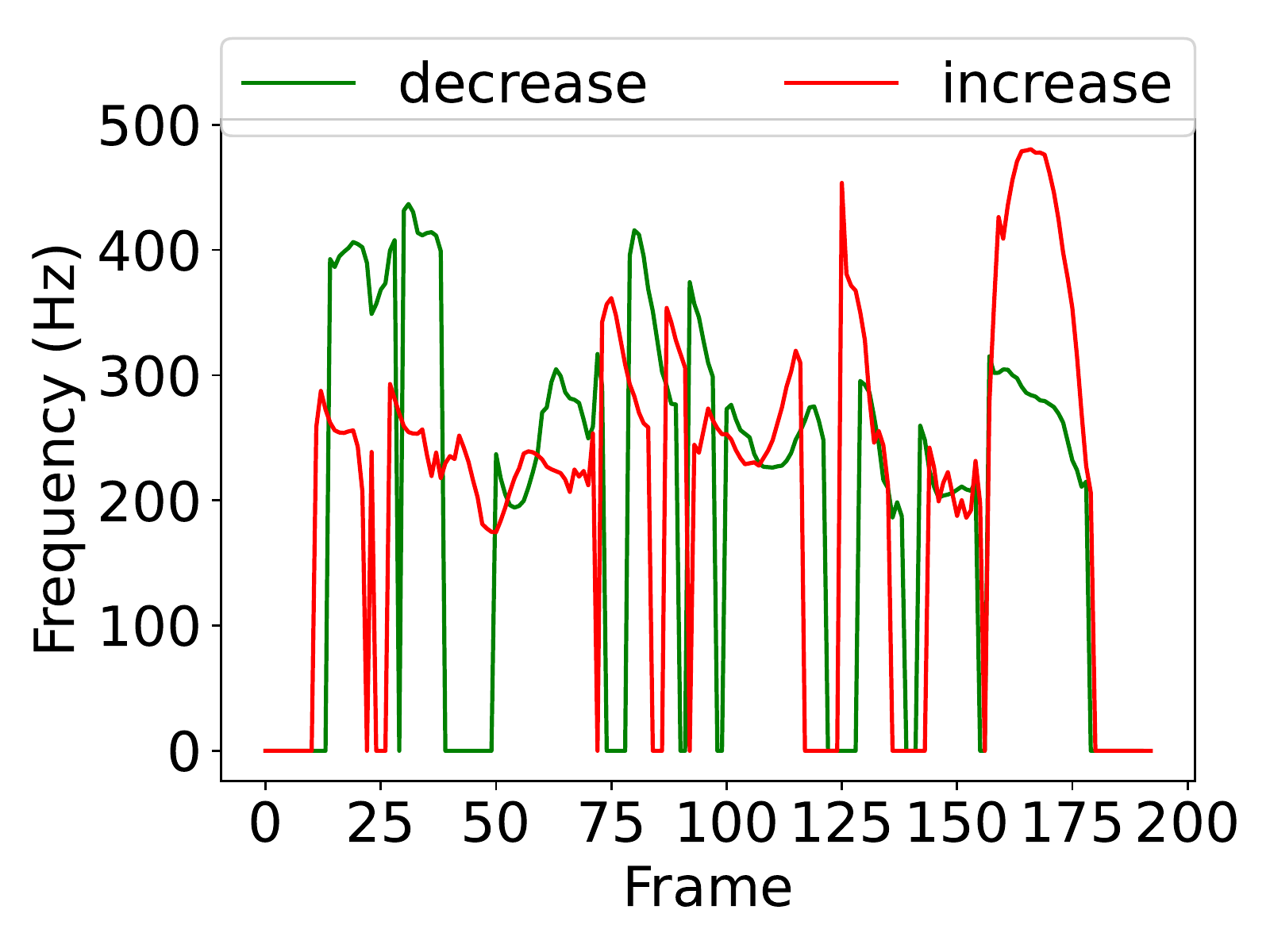}}
    \label{1b}\\
	  \subfloat[sadness]{
        \includegraphics[width=0.45\linewidth]{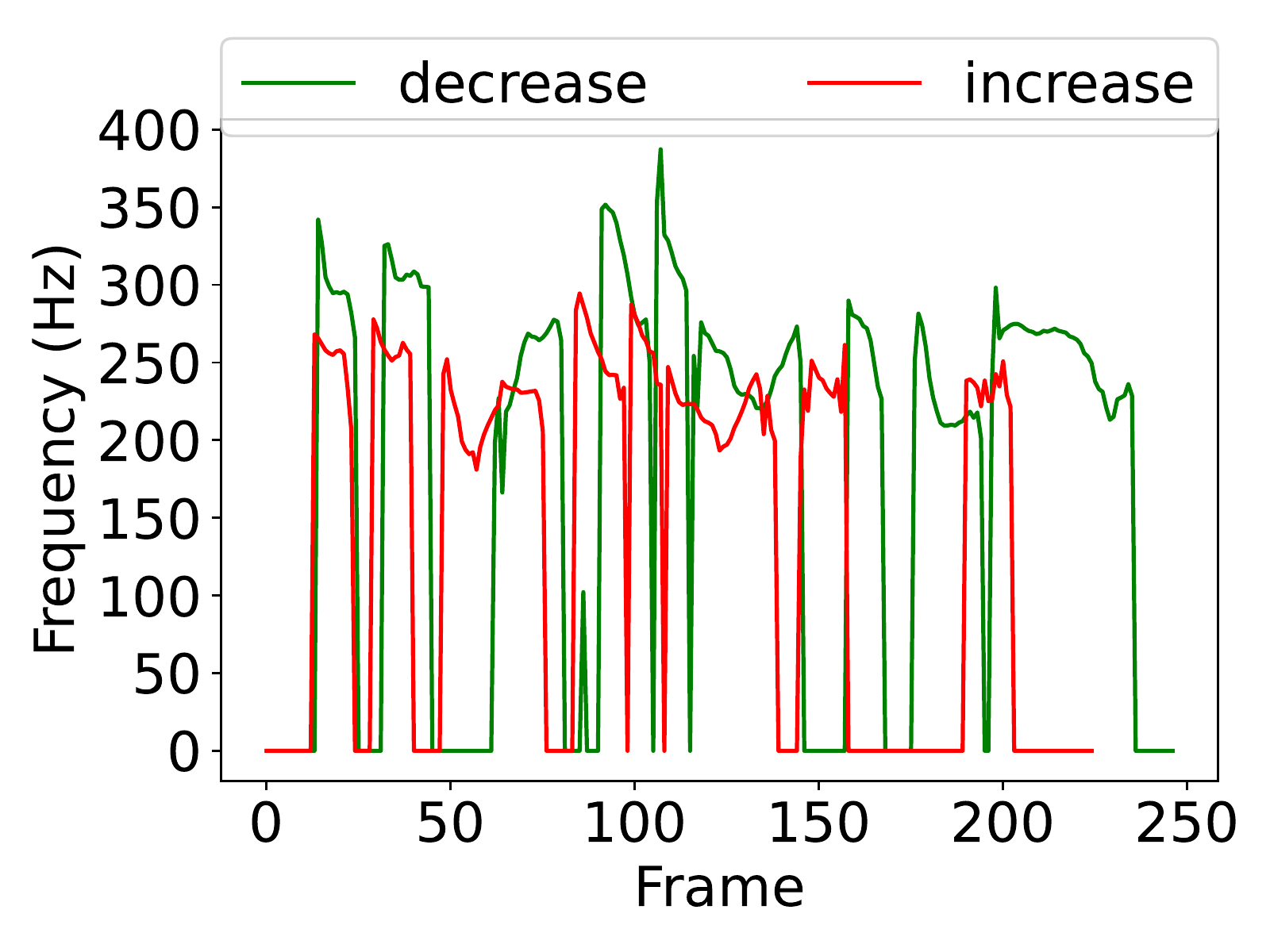}}
    \label{1c}\hfill
	  \subfloat[surprise]{
        \includegraphics[width=0.45\linewidth]{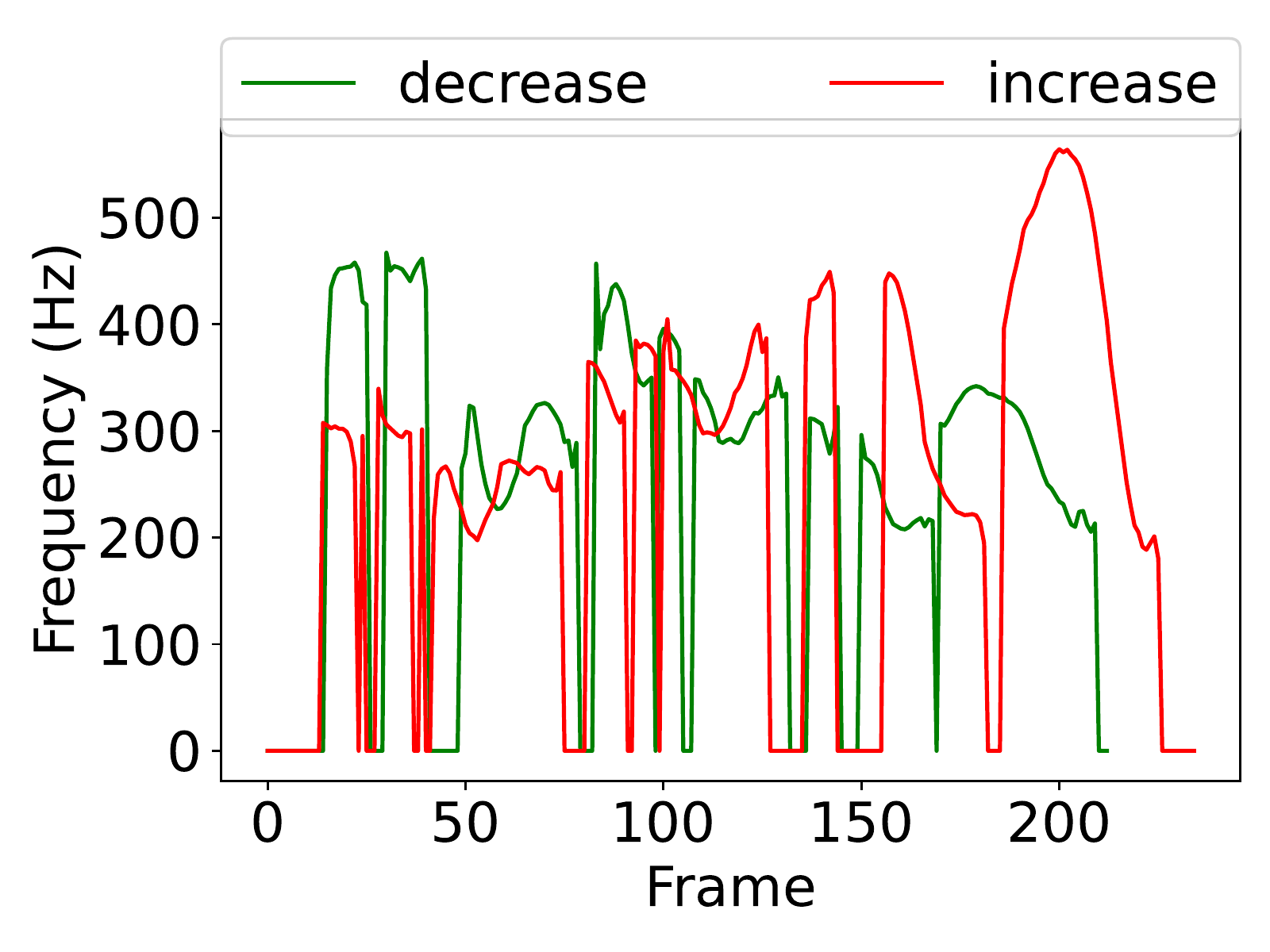}}
     \label{1d} 
     \subfloat[fear]{
        \includegraphics[width=0.45\linewidth]{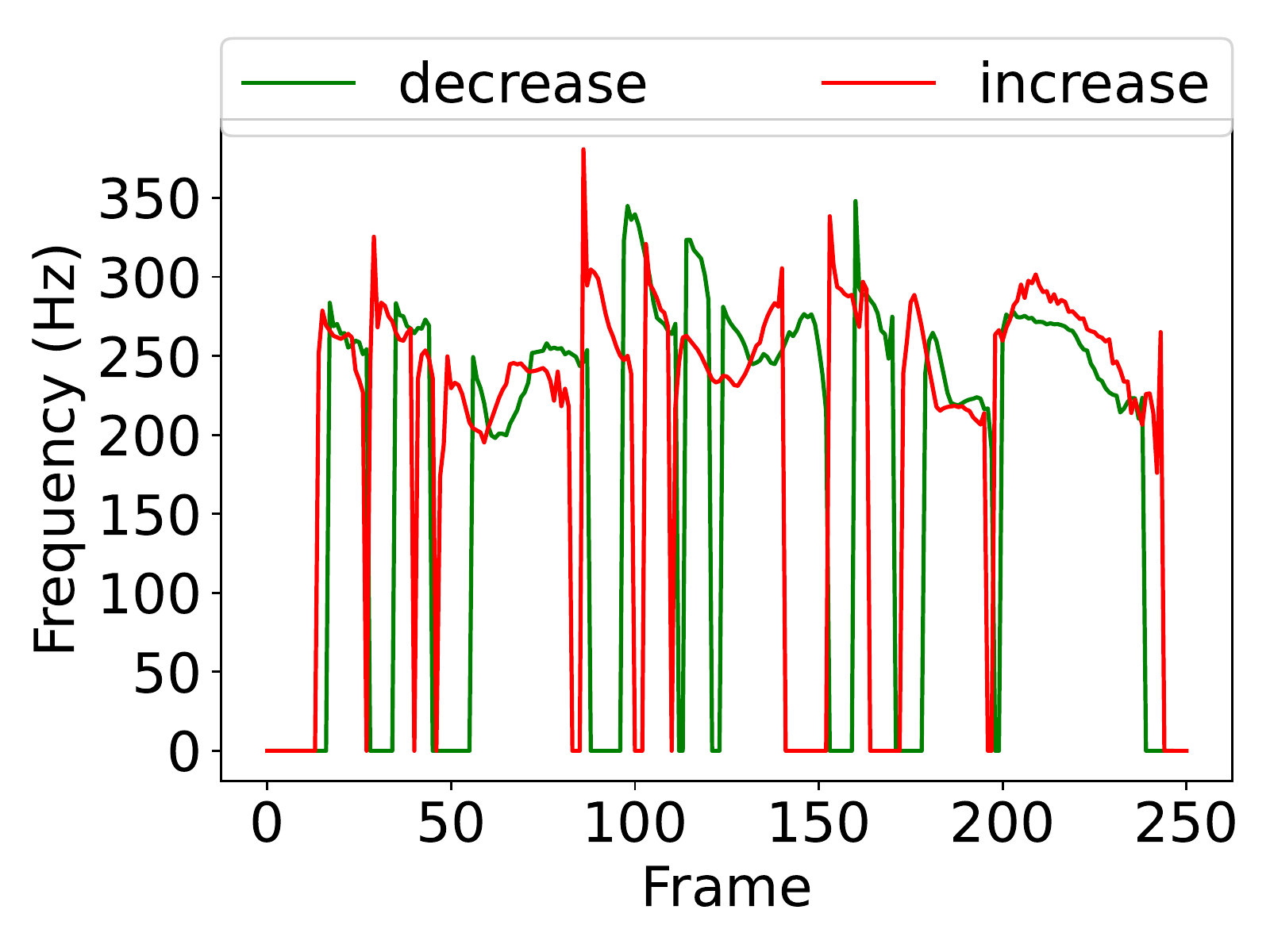}}
    \label{1e}\hfill
	  \subfloat[disgust]{
        \includegraphics[width=0.45\linewidth]{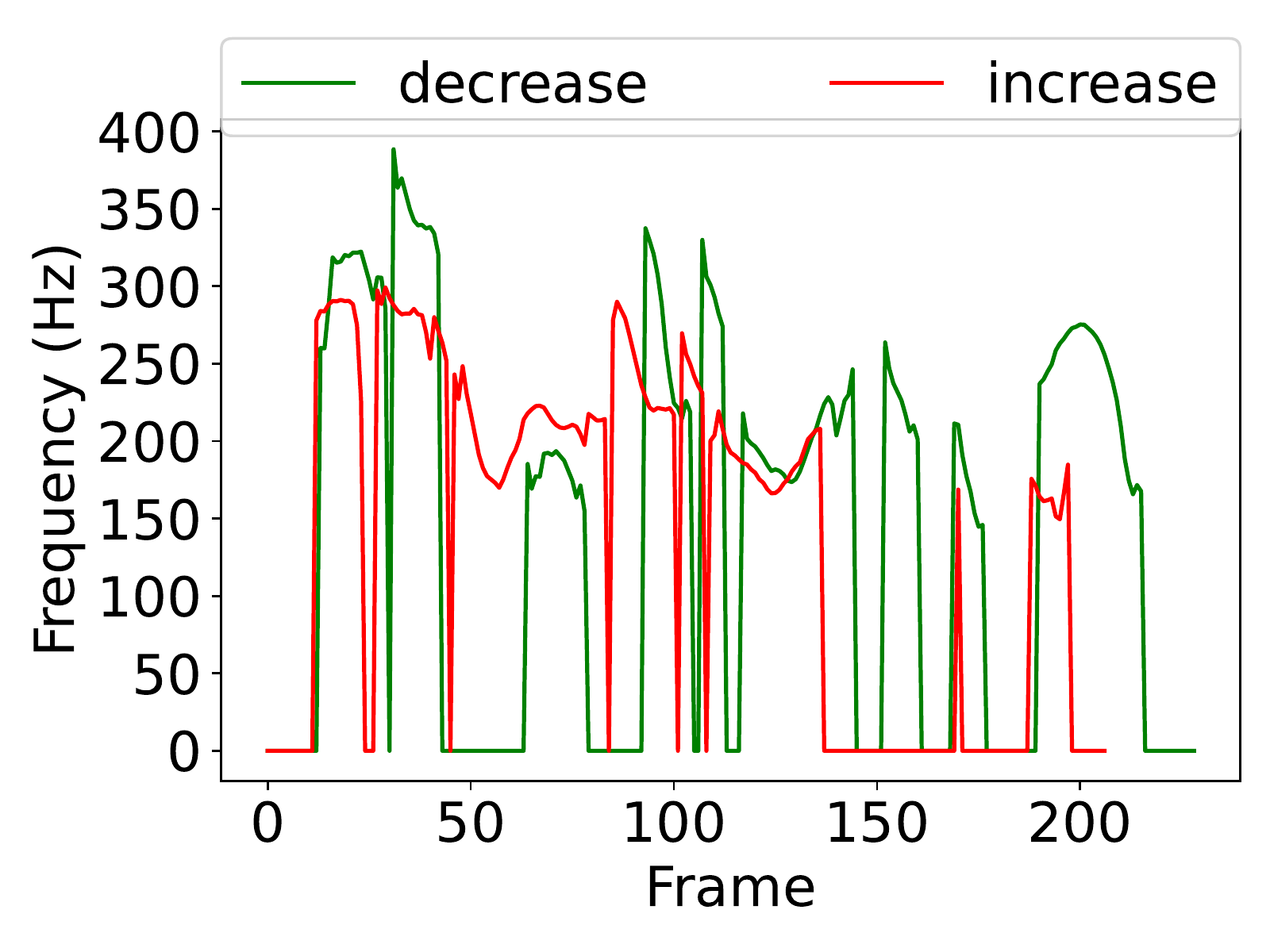}}
     \label{1f} 
	  \caption{F0 curves of synthesized samples with gradually increased or decreased strengths of different emotion strengths}
	  \label{fig:ctl_ud} 
\end{figure}

Further analysis for the ability of fine-grained emotion control of MsEmoTTS is shown in Fig.~\ref{fig:ctl_ud}, in which the local strengths gradually increase or decrease within an utterance. Specifically, we assign values from $0$ gradually increasing to $1$, or from $1$ gradually decreasing to $0$ for the syllable emotion strengths. As can be seen from this figure, F0 or duration or both of them change gradually in the utterances, presenting overall increases or decreases with the local strengths changing, demonstrating the ability of the proposed model to achieve flexible fine-grained emotion control. Note that in the ``disgust'' emotion, the larger strengths result in the quicker change of pitch and duration, and lower strengths have a more stable speaking style. Although different emotion categories have different patterns of expression for high emotion strengths, all the results accord with the way of emotion itself. We suggest the readers listen to our online demos~\footnote{Audio demos can be found at \url{https://leiyi420.github.io/MsEmoTTS/}}.

\section{Component analysis}
\label{exp_hierarchical}
In Section~\ref{sc:exp_results}, MsEmoTTS shows good performance on emotional speech synthesis based on emotion transfer, prediction, and control. In this section, the effectiveness of each proposed module, i.e., GM, UM, and LM, will be analyzed.

\begin{table}[htb]
\centering
\topcaption{Comparison of different text emotion classifier. DNN-based is a classifier of DNN-based network jointly trained with the acoustic model, and BERT-based is the pre-trained emotion classifier utilized in this work.}
\label{tab:emo-cls}
\setlength{\tabcolsep}{2mm}
\begin{tabular}{cccc}
\toprule
Method & Overall accuracy (\%)  & F1 Macro (\%) & Kappa value\\ \midrule
DNN-based & 33  & 34 & 0.125  \\ 
BERT-based   & 55  & 57 & 0.475 \\ \bottomrule
\end{tabular}
\end{table}

\begin{table*}[tp]
\centering
\topcaption{Ablation study for the UM on the emotion transfer speech synthesis task. w and w/o means with or without the UM module respectively, and $p$ denotes the $p$-value of a $t$-test between two models. CMOS and A/B preference tests are performed according to the emotion similarity. Positive CMOS value means that the model with UM is better than the model without UM.}

\label{tab:cmos-ab-ul}
\setlength{\tabcolsep}{3mm}
\begin{tabular}{c|c|cccc|c|c|cccc}
\hline
    \multicolumn{1}{c|}{\multirow{2}{*}{Parallel}}&
    \multicolumn{1}{c|}{\multirow{2}{*}{CMOS}} &
    \multicolumn{4}{c|}{Preference(\%)} &
    \multicolumn{1}{c|}{\multirow{2}{*}{Nonparallel}} &
    \multicolumn{1}{c|}{\multirow{2}{*}{CMOS}} &
    \multicolumn{4}{c}{Preference(\%)} \\
    \cline{3-6} \cline{9-12}
    & & w/o  & Neutral & w/  & $p$ & & & w/o  & Neutral & w/  & $p$ \\ \midrule
    Happiness  & 0.562 & 16.7 & 33.3 & \textbf{50.0} &  0.026 & Happiness  & 0.082 & 27.2 & 35.8 & \textbf{37.0} & 0.043 \\
    Anger  & 0.593 & 18.3 & 25.0 & \textbf{56.7} &  0.013 & Anger  & 0.512 & 13.3 & 25.9 & \textbf{60.8} & $<$ 0.001\\
    Sadness  & 0.333 & 20.8 & 33.2 & \textbf{46.0} &  0.034 & Sadness  & 0.078 & 28.6 & 34.7 & \textbf{36.7} & 0.041\\
    Surprise  & 0.250 & 25.0 & 29.2 & \textbf{45.8} &  0.037 & Surprise  & 0.375 & 19.4 & 38.8 & \textbf{41.8} & 0.039\\
    Fear  & 0.608 & 12.5 & 25.0 & \textbf{62.5} & $<$ 0.001 & Fear  & 0.358 & 28.6 & 31.4 & \textbf{40.0} & 0.042\\
    Disgust  & 0.650 & 16.7 & 20.8 & \textbf{62.5} &  $<$ 0.001 & Disgust  & 0.458 & 18.5 & 33.9 & \textbf{47.6} & 0.024\\ \midrule
    Average   &  0.499 &  18.3 &  27.8 & \textbf{53.9} &  0.003 & Average & 0.311 & 22.6 &  33.4 & \textbf{44.0} & 0.034 \\ \bottomrule
\end{tabular}
\end{table*}

\subsection{Global-level emotion presenting module (GM)}
\label{exp_global}

The GM is the key module to provide the emotion category, and no other module exists to provide this information, making the GM indispensable in the proposed method for the emotional speech synthesis. Therefore, we will not conduct an experiment to prove the effectiveness of the GM. Instead, we evaluate the performance of our emotion classifier and the proposed \textit{soft} emotion embedding. 

To evaluate the performance of the proposed emotion classifier on the emotion classification task, a comparison between the BERT-based classifier and a DNN-based classifier that is jointly trained with the acoustic model, which is adopted in our preliminary work~\cite{lei2021fine}, is conducted. Table~\ref{tab:emo-cls} shows the classification results of the two methods. As can be seen, the proposed BERT-based emotion classifier outperforms the DNN-based method in terms of all evaluation metrics, indicating the good design of the proposed emotion classifier.

\begin{table}[t]
\vspace{10pt}
\centering
\topcaption{Subjective comparison between the proposed method and a variant model, referred to as P-h, in which the \textit{soft} emotion embedding method is replaced with the \textit{hard} emotion embedding method. MOS results on the text-based emotional speech synthesis task are reported with 95\% confidence interval. Higher MOS means better performance.}
\label{tab:hard}
\setlength{\tabcolsep}{5.5mm}
\begin{tabular}{c|c|c}
\toprule
MOS       & P-h & MsEmoTTS (Proposed) \\ \midrule
Happiness & 3.65 $\pm$ 0.148 & \textbf{4.19 $\pm$ 0.084} \\
Anger     & 3.73 $\pm$ 0.126 & \textbf{4.04 $\pm$ 0.128} \\
Sadness   & 3.38 $\pm$ 0.167 & \textbf{3.95 $\pm$ 0.156} \\
Surprise  & 3.61 $\pm$ 0.149 & \textbf{4.15 $\pm$ 0.117} \\
Fear      & 3.50 $\pm$ 0.095 & \textbf{4.05 $\pm$ 0.075}  \\
Disgust   & 3.34 $\pm$ 0.183 & \textbf{3.75 $\pm$ 0.156} \\ \midrule
Average   & 3.54 $\pm$ 0.145 & \textbf{4.02 $\pm$ 0.119} \\ \bottomrule
\end{tabular}
\end{table}

Considering the purpose of automatically synthesizing emotional speech, we would like to show the superiority of the proposed \textit{soft} emotion embedding to the \textit{hard} emotion embedding (see Section~\ref{sc:method_GM}), rather than focusing on the text emotion classification.

The comparison of the \textit{hard} emotion embedding-based method (``P-h'' in the table) and the \textit{soft} emotion embedding-based (proposed) method is shown in Table~\ref{tab:hard}, in which the MOS test is performed on the task of emotional speech synthesis by predicting the emotion from the input text. The results of the soft-based method are the same as the results of the proposed method in Table~\ref{tab:mos}. Here, for the convenience of comparison, these results are also listed in Table~\ref{tab:hard}. As shown in this table, in terms of all emotion categories, the proposed \textit{soft} emotion embedding-based method significantly outperforms the \textit{hard}-based method. On the average MOS, the proposed \textit{soft}-based method is 13.6\% relatively higher than the \textit{hard}-based method. These results indicate the effectiveness of the proposed \textit{soft} emotion embedding method.

\begin{figure}[ht]
        \centering
        \includegraphics[width=1.0\linewidth]{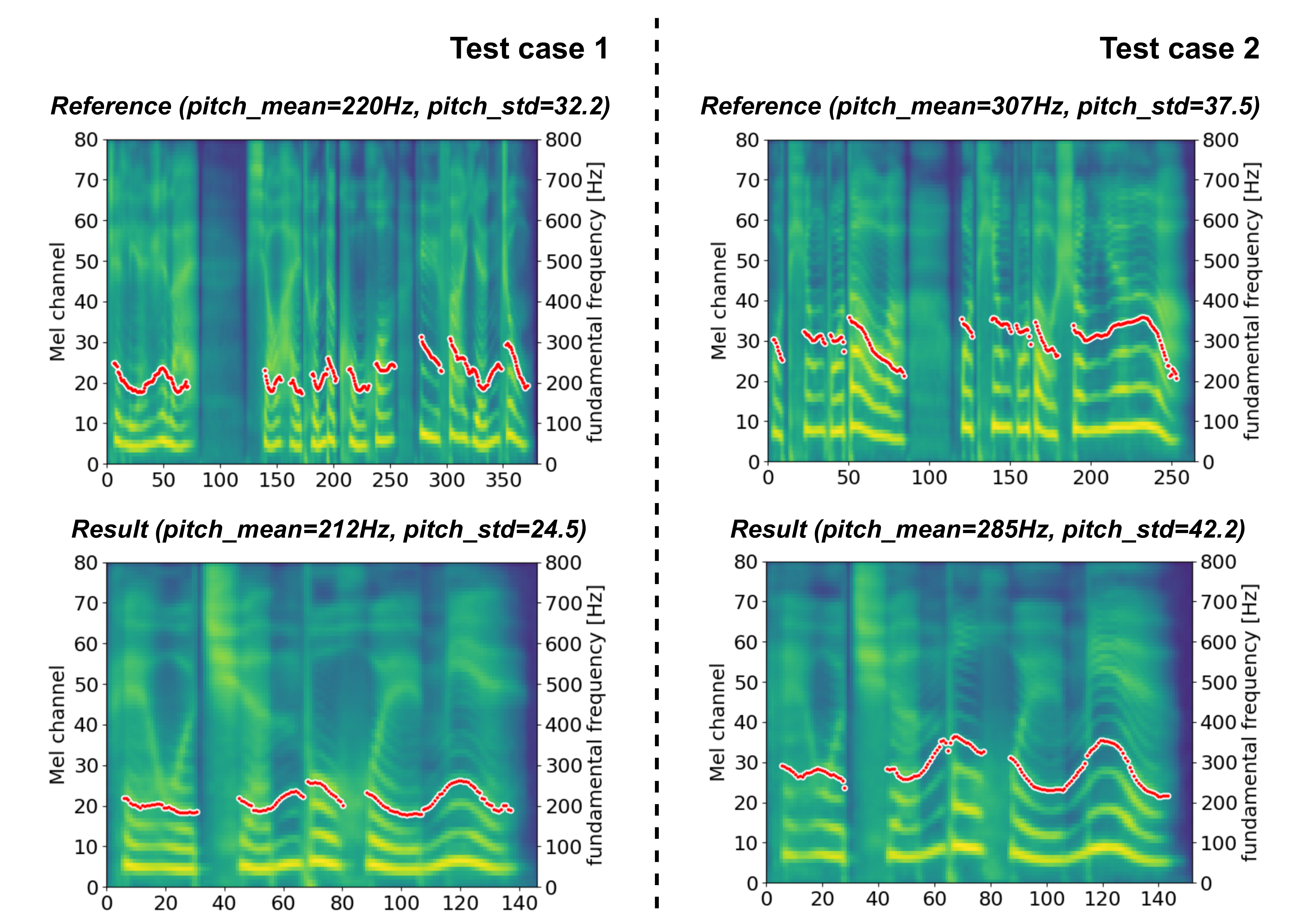}
        \caption{Mel-spectrograms and F0 contours of generated speech with different utterance-level reference signals that sharing the same global-level emotion category and local-level emotion strengths.}
         \vspace{-10pt}
        \label{fig:um-mel_pitch}
\end{figure}

\subsection{Utterance-level emotion presenting module (UM)}
\label{exp_utterance}

To analyze the effect of UM, an ablation study is performed in this section. To be specific, we compare the proposed method (with UM) and a variant of the proposed method (without UM) on the emotion transfer speech synthesis task. The comparison results are shown in Table~\ref{tab:cmos-ab-ul}, in which both CMOS and A/B preference tests are reported for parallel and non-parallel emotion transfer speech synthesis tasks. In this table, the positive CMOS value means the proposed method with UM is better than that without UM. As shown in this table, the proposed method significantly outperforms the model without UM both in CMOS test and A/B preference test in terms of all emotion categories, which indicates the effectiveness of modeling the utterance-level variation for the emotional speech synthesis.

To further analyze what the UM learns, two different utterance-level reference signals are taken to synthesize speech with the same global-level emotion category and local-level emotion strengths. Fig~\ref{fig:um-mel_pitch} intuitively shows the synthesized results with different utterance-level reference signals, in which the mel-spectrograms and f0 contours are presented. It can be found that the pitch variation of synthesized speech is closely related to the reference audio's pitch. Specifically, the second generated audio has a higher and more varied pitch than the first generated audio in the utterance level, which is consistent with the corresponding reference signals. These results indicate that UM is able to learn the trend of intonation within the sentences or the general utterance-level pattern of the prosody.

\subsection{Local-level emotion presenting module (LM)}
\label{exp_local}

To demonstrate the effectiveness of the local emotion strengths on the emotional speech synthesis, we replace the syllable-level emotion strengths with the sentence-level emotion strengths, in which way only the global emotion strength can be provided rather than the fine-grained local level. Besides, an alternative speech unit used in our preliminary work~\cite{lei2021fine}, i.e., phoneme, for the local emotion strength is also compared in this section. The comparison is conducted with the parallel reference audio-based emotional speech synthesis task, which allows us to evaluate the performance objectively. To avoid the influence from the utterance variance prediction to the analysis, in this comparison, the variation feature is not adopted. 

\begin{table}[htb]
\centering
\topcaption{Comparison of models with different levels to present local emotion strengths. MCD is reported on the parallel emotion transfer speech synthesis task. The lower MCD means better performance.}
\label{tab:mcd-local}
\setlength{\tabcolsep}{12mm}
\begin{tabular}{cc}
\toprule
Level & MCD (dB) \\ \midrule
Phoneme-level & 4.16  \\ 
Syllable-level    & \textbf{4.11} \\ 
Sentence-level       & 4.65 \\ \bottomrule
\end{tabular}
\end{table}

The results are shown in Table~\ref{tab:mcd-local}, in which \textit{Syllable-level} is the proposed method, \textit{Phoneme-level} is another compared local strength method in our preliminary work~\cite{lei2021fine}, and \textit{Sentence-level} does not take the local emotion strength into consideration. As shown in this table, both \textit{Phoneme-level} and \textit{Syllable-level} based methods are significantly better than the \textit{Sentence-level} based method. Specifically, MCD scores of \textit{Phoneme-level} and \textit{Syllable-level} based methods are 10.5\% and 11.6\%, respectively, relatively lower than that achieved by the \textit{Sentence-level} based method, indicating that the local emotion strengths are effective. 

Compared with the \textit{Phoneme-level} based method, the \textit{Syllable-level} based method shows slight better performance. The possible reason is that in Chinese -- a character-based language where each character is pronounced as a toned syllable, the variation of pitch is saliently reflected in the syllable, making the syllable more suitable for the local-level modeling in Chinese.

\section{Discussion}
\label{sc:discussion}

With all three modules, i.e., GM, UM, and LM, the proposed method can synthesize emotional speech either by transferring from reference speech or by predicting from the input text. Besides, the global emotion category in GM and the local emotion strength in LM can be manually defined, in which way the proposed method can synthesize emotional speech as expected.

In the current method, both the emotion classifier and emotion strength extractor are pre-trained for better performance, which can benefit from more available data. However, this step-wise method also increases the complexity of model training. For future work, jointly learning the classifier under a multi-task framework is an alternative way to learn the global emotion representations within a unified acoustic model. As for the local emotional strengths, it is worth exploring how to model the strengths with neural networks in an unsupervised manner, which benefits to simplify both the training and generation process.

It is worth noting that although the references for GM and UM are independent of each other in the practice, our extensional experiments show that when the references for GM and UM are from different emotion categories, the synthesized speech somehow sounds weird, which sounds like an emotion between the references for GM and UM. This is because the variation within an utterance is related to the emotion category.

\section{Conclusion}
\label{Conclusion}
Inspired by the hierarchical nature of prosody, a multi-scale model for emotional speech synthesis, called MsEmoTTS, is proposed in this paper. MsEmoTTS allows modeling the emotion from different levels and performing the emotional speech synthesis in different ways. Extensive experiments demonstrate that MsEmoTTS achieves good performances on emotional speech synthesizing by transferring the emotion from reference audio or predicting the emotion only from input text, and also show that MsEmoTTS can be controlled manually to synthesize emotional speech as expected.

\ifCLASSOPTIONcaptionsoff
  \newpage
\fi



%
\bibliographystyle{IEEEtran}  
\bibliography{IEEEabrv,main}


%







\end{document}